\documentclass[sn-basic]{sn-jnl}

\usepackage{graphicx}%
\usepackage{multirow}%
\usepackage{amsmath,amssymb,amsfonts}%
\usepackage{amsthm}%
\usepackage{mathrsfs}%
\usepackage{textcomp}%
\usepackage{manyfoot}%
\usepackage{booktabs}%
\usepackage{algorithm}%
\usepackage{algorithmicx}%
\usepackage{algpseudocode}%
\usepackage{listings}%
\newcommand{\tcr}[1]{#1}
\newcommand{\cellcolor}[1]{}

\theoremstyle{thmstyleone}%
\newtheorem{theorem}{Theorem}
\newtheorem{proposition}[theorem]{Proposition}%

\theoremstyle{thmstyletwo}%
\newtheorem{remark}{Remark}%
\newtheorem{lemma}[theorem]{Lemma}
\newtheorem{corollary}[theorem]{Corollary}
\theoremstyle{thmstylethree}%
\newtheorem{definition}{Definition}%

\begin{document}

\title{A Bayes-Motivated Quadratic-Form Test for High-Dimensional Mean Testing}

\author[1]{\fnm{Daojiang} \sur{He}}\email{djhe@ahnu.edu.cn}

\author[1]{\fnm{Suren} \sur{Xu}}\email{2221012661@ahnu.edu.cn}

\author*[2]{\fnm{Jing} \sur{Zhou}}\email{jing.zhou@manchester.ac.uk}

\affil[1]{Department of Statistics, Anhui Normal University, China}

\affil*[2]{Department of Mathematics, University of Manchester}


\abstract{
We propose a two-sample mean test based on the Bayes factor with non-informative priors, specifically designed for scenarios where \tcr{the dimension $p$ grows with the sample size $n$ with a linear rate $p/n \to c_1 \in (0, \infty)$}. We establish the asymptotic normality of the test statistic and the asymptotic power. Through extensive simulations, we demonstrate that the proposed test performs competitively \tcr{against several exisiting methods}, particularly when the \tcr{the marginal variances of the individual features} are heterogeneous and when the sample size is small. 
Furthermore, our test remains robust under distribution misspecification. The proposed method not only effectively detects both sparse and non-sparse differences in mean vectors but also maintains a well-controlled type I error rate, even in small-sample scenarios. We also demonstrate the performance of our proposed test using the \tcr{small round blue cell tumors (SRBCT)} dataset.}




\maketitle

\section{Introduction}
\label{sec:intro}

In modern statistical and machine learning applications, determining whether samples come from the same population is a fundamental issue for exploring data structure and defining problems. When merging information from different sources, it is especially interesting to know if the sources are roughly homogeneous. 
In higher-dimensional or multivariate settings, directly comparing distributions presents additional challenges, especially when sample sizes are small. Small sample sizes can lead to greater variability in test results, making it harder to draw reliable conclusions. A simple solution is to test whether the moments of the two samples match, which is well-explored in the classic small $p$ setting known as Hotelling's $T^2$ test in \cite{hotelling1931generalization}. The test typically assumes that the two samples $X_{1,1},\ldots,X_{1,n_1} \stackrel{iid}{\sim} N_p(\mu_1, \Sigma)$ and $X_{2, 1},\ldots, X_{2, n_2} \stackrel{iid}{\sim} N_{p}({\mu_2},{\Sigma})$ and we are interested in testing 
\begin{equation}\label{h0h1}
	H_0:\mu_{1}=\mu_{2} = \mu \mbox{ versus } H_1:\mu_{1}\neq \mu_{2}.
\end{equation} 
Hotelling's $T^2$ test becomes less appealing when $p$ is nonnegligible, sometimes exceeding the sample sizes. One major issue is that the sample covariance matrix is not a consistent estimator of the population one when $p$ diverges. Further, the pooled sample covariance matrix 
\begin{eqnarray}\label{eq: pooled covariance matrix}
	S_n =\frac{1}{n -2}\left\{\sum_{i=1}^{n_{1}}({X}_{1,i}-\bar{X}_1)(X_{1,i}-\bar X_1)^\top +\sum_{i'=1}^{n_{2}}(X_{2,i'}-\bar X_2)(X_{2,i'}-\bar X_2)^\top\right\},
\end{eqnarray}
when $p$ exceeds the sample size $n = n_1+n_2$, is ill-defined. An early attempt is \cite{bai1996effect}, which constructed a test statistic by subtracting the diagonal of the sample covariance matrix following 
\begin{equation}\label{eq: BS test stat}
	T_{\mathrm{BS}}=\frac{(n_1^{-1} + n_2^{-1})^{-1}(\bar X_1 - \bar X_2)^\top (\bar X_1 - \bar X_2) - \mbox{tr}(S_n)}{[2n(n+1)(n-1)^{-1}(n+2)^{-1} \{ \mbox{tr}(S_n^2) - n^{-1} (\mbox{tr}S_n)^2\} ]^{1/2}}.
\end{equation}
After noting that the cross product in \eqref{eq: BS test stat} causes estimation problems, \cite{chen2010two} proposed a new test statistic following 
\begin{equation}\label{eq: CQ test statistic}
	T_{\rm CQ} = \frac{\sum_{i \neq i'}^{n_1} X_{1i}^\top X_{1i'}}{n_1(n_1 - 1)} + \frac{\sum_{i \neq i'}^{n_2} X_{2i}^\top X_{2i'}}{n_2(n_2 - 1)} -  \frac{2\sum_{i=1}^{n_1}\sum_{i' =1}^{n_2} X_{1i}^\top X_{2i'}}{n_1n_2}.
\end{equation}
The proposed test statistic is shown to achieve comparable asymptotic power as \eqref{eq: BS test stat}. To address the components of the multivariate Gaussian distribution having different variances, \cite{srivastava2008test} used the \tcr{diagonal elements of the covariance matrix, that is, the marginal variances of the individual features,} to improve \eqref{eq: BS test stat}, which led to
\begin{equation}\label{eq: SD test stat}
	T_{\mathrm{SD}}=\frac{(n_1^{-1} + n_2^{-1})^{-1}(\bar X_1 - \bar X_2)^\top D_{S_n}^{-1} (\bar X_1 - \bar X_2) - (n-4)^{-1}(n-2)p}{\sqrt{2\left[\mathrm{tr}(R^2)-\frac{p^2}{n}\right]c_{p,n}}},
\end{equation}
where $D_{S_n}$ is the diagonal of $S_n$,\;$c_{p,n}=1+\mathrm{tr}(R^2)p^{-3/2}$ and $R=D_{S_n}^{-\frac{1}{2}}S_n D_{S_n}^{-\frac{1}{2}}$. Importantly, this substitution enhances the stability in high-dimensional settings when the full covariance matrix is difficult to estimate consistently. Incorporating the precision matrix in such cases may lead to excessive variability or singularity.

The above tests are constructed using the $L_2$-norm of the two sample means, with the most recent work to our knowledge \cite{yang2024new} addressing such tests by proposing a new estimator for the ill-conditioned covariance matrix. When the difference between the two sample mean vectors is sparse, test statistics based on $L_\infty$-norm \cite{tony2014two,xue2020} consider the maximum difference of the two sample means. Such tests are less affected by the large proportion of zeros of the difference of the mean vectors as opposed to the $L_2$-norm-based tests. 

An alternative method is known as ``random projection", which involves the transformation of high-dimensional data into lower-dimensional data through projection. For example, \cite{lopes2011more, srivastava2016raptt} used random projection to obtain the weight of the mixture of high-dimensional Gaussian random variables; \cite{guhaniyogi2015bayesian} used random projection to propose a Bayesian compression regression. The readers are referred to \cite{huang2022overview} for a thorough overview of the topic.

However, testing two sample means for large $p$ is unexplored in the Bayesian statistics framework. Recently, some work has considered using the Bayes factor to construct test statistics. For instance, \cite{zoh2018powerful} considered random projection to maximize the power of the test based on the Bayesian factor. The high dimensional data are projected to low dimensions such that the Bayes factor is applicable. An important issue of such a projection is that the test is not invariant under the scaling of the data. To address this, \cite{thulin2014high,chen2024bayesian} considered splitting the high dimensional data into clusters such that the test statistics are well-defined on the clusters. In a similar spirit to the frequentist' two-sample tests in high dimensions, \cite{jiang2022two} proposed a test based on the posterior Bayes factor \cite{aitkin1991posterior}. By setting an inverse Wishart prior with degrees of freedom $m$ for the population covariance matrix of the two samples $\Sigma$, the test statistic can be equivalently expressed as 
\begin{equation}\label{eq: jiang xu test}
	T_{\mathrm{PB}}=\frac{(n_1^{-1} + n_2^{-1})^{-1} (\bar X_1 - \bar X_2)^\top B(\bar X_1 - \bar X_2)- \mbox{tr}(BS_n)}{\sqrt{2\left[\mbox{tr}(BS_n)^2-\frac{1}{n-2}(\mbox{tr}(BS_n))^2\right]}},
\end{equation}
where $B=2(m+2n)\left(2A_n + k' I _p\right)^{-1}-(m+n)\left(A_n+ k' I_p\right)^{-1}$,\;$A_n=(n-2)S_n$ and $ k'$ is a hyperparameter controlling the variance of the prior of $\Sigma$. Their proposed test does not require random projection and has been shown to achieve better numerical performance than \cite{zoh2018powerful,chen2010two,srivastava2008test}. Additionally, \cite{lee2024bayesian} proposed a maximum pairwise Bayes factor approach by testing the largest difference of the mean vectors, which may have power loss when the difference is dense.

We propose a two-sample test motivated by a Bayesian formulation and based on the Bayes factor for high-dimensional settings in which $p$ is large. In particular, we focus on scenarios in which the multivariate Gaussian distribution exhibits different variances across components, as discussed in \cite{srivastava2008test}. Importantly, by following \cite{srivastava2008test} and using only the diagonal elements $D_{S_n}$ in place of the full sample covariance matrix $S_n$, we enhance the reliability and stability of the test in high-dimensional settings, particularly when the sample size is small and consistent estimation of the full covariance structure is infeasible. Incorporating full covariance information in such regimes may introduce substantial variability or even lead to singularity, potentially compromising control of type I error and limiting the overall applicability of the test. \tcr{We clarify that the Bayesian formulation in the paper is used as the derivational starting point for identifying a quadratic-form statistic, and the final procedure is then analyzed through the statistics' large-sample properties, see \cite{jiang2022two}, for a similar construction.} We establish the asymptotic normality of the \tcr{resulting} test statistic under the null hypothesis $H_0$ and derive the asymptotic power. The proposed test demonstrates stable performance for both sparse and non-sparse differences in mean vectors. Notably, by incorporating bias correction coefficients, the proposed test achieves controlled size in small samples, improving upon existing competing tests, including \tcr{the BS, CQ, SD, and PB procedures, utilizing the test statistics defined in \eqref{eq: BS test stat}, \eqref{eq: CQ test statistic}, \eqref{eq: SD test stat}, and \eqref{eq: jiang xu test}, respectively. We also include the random-projection-based approach from \cite{zoh2018powerful} as a benchmark, denoted RM hereafter.}  The proposed test exhibits robust, competitive empirical performance under violation of Gaussianity.

The paper is organized as follows. In Section~\ref{sec: test stat based on bayes factor}, we present the Bayes factor $\mbox{BF}_{10}$, and derive the null distribution of the Bayes factor and the asymptotic power function. 
Section~\ref{sec: Test statistic} constructs the test statistic based on a component of the Bayes factor and proposes ratio consistent estimators of the unknown quantities in the test statistic. Further, we propose correction coefficients to adjust the proposed test statistic for small samples in Section~\ref{small sample correction}; such a correction is important to guarantee a controlled empirical size of the test. In Section \ref{sec: numerical}, we carry out a simulation study to assess the performance of the proposed test. In Section \ref{sec: real data analysis}, we utilize the proposed test to analyze a real dataset. In Section \ref{sec:conc}, we give some discussions about this article.

\section{Two sample mean test based on the Bayes factor}\label{sec: test stat based on bayes factor}
Assume having the following two samples
\[X_{1,1},\ldots,X_{1,n_1} \stackrel{iid}{\sim} N_p(\mu_1, \Sigma) \mbox{ and } X_{2, 1},\ldots, X_{2, n_2} \stackrel{iid}{\sim} N_{p}({\mu_2},{\Sigma}),\]
we are interested in testing \eqref{h0h1}, that is, if they share a common mean vector. The Gaussian assumption is standard in the existing literature. It is important for obtaining a closed-form expression of the Bayes posterior distributions in \eqref{eq: bayes factor} for constructing the test statistic. \tcr{The theoretical results in this paper are established under the Gaussian assumption, which is common in the literature on Bayes factor-based test statistics; see, for example, \cite{wang2021bayesian, jiang2022two}. Extending the proposed framework to spherically symmetric errors would substantially complicate both the derivation of the test statistic and the analysis of its asymptotic distribution and power; see, for example, \cite{wang2024bayesian}. We therefore leave this extension for future work. 
}
Further, we will show in Section~\ref{sec: numerical} that the constructed test can be applied to sub-Gaussian distributions as empirical evidence of robustness. 

Our test statistic is based on the closed-form expression of the Bayes factor, defined as
\begin{eqnarray}\label{eq: bayes factor definition}
	\lefteqn{
    \mbox{BF}_{10}=\frac{ f(X_{1,1}, \ldots, X_{1, n_1}, X_{2, 1}, \ldots, X_{2, n_2} \mid H_{1})}{f(X_{1,1}, \ldots, X_{1, n_1}, X_{2, 1}, \ldots, X_{2, n_2} \mid H_{0})}}&& \nonumber\\
	&=&\frac{\int l(X_{1,1}, \ldots, X_{1, n_1}, X_{2, 1}, \ldots, X_{2, n_2} |\Theta)\pi_1({\Theta |H_{1}}) d\Theta}{\int l(X_{1,1}, \ldots, X_{1, n_1}, X_{2, 1}, \ldots, X_{2, n_2} |\Theta)\pi_0(\Theta|H_{0}) d\Theta},
\end{eqnarray}
where $\pi(\Theta|H_0)$ and $\pi(\Theta|H_1)$ are the specified prior distributions for the unknown parameters $\Theta = (\mu_1, \mu_2, \Sigma)$ under $H_0$ and $H_1$, respectively. Following \cite{jiang2022two}, under the null 
\begin{equation}\label{eq: null hypothesis}
	H_0: \mu_1 = \mu_2 = \mu, 
\end{equation}
the prior distributions are chosen to be
\begin{equation}\label{eq: prior null}
	\pi_0(\mu)=1,\;\pi_0(\Sigma)\sim W_p^{-1}(m_0, V),
\end{equation}
and
\begin{equation}\label{eq: prior alternative}
	\pi_1(\mu_1)=\pi_1(\mu_2)=1,\;\pi_1(\Sigma)\sim W_p^{-1}(m_1, V),
\end{equation}
where $W_p^{-1}(m, V)$ is an inverse Wishart distribution with degrees of freedom $m$ and a positive definite matrix $V$. Following \cite{zoh2018powerful}, we set \tcr{$P(H_0)=P(H_1)=0.5$, that is, the two competing hypotheses are assigned equal prior probabilities.} 
\tcr{Our goal is not to use the full Bayes factor as the final testing rule, but to extract from it a stable high-dimensional quadratic-form component. In that reduced statistic, the prior mainly enters through $k'$ in $V=k'I_p$, so the influence of the prior is much more limited than in a fully Bayesian analysis based on the complete Bayes factor.} Importantly, to test if the proposed Bayes factor test statistic outperforms the competitors, especially the one in \eqref{eq: jiang xu test}, we are interested in a test statistic which has stable performance under non-informative priors but with less restriction on the variance of the covariance matrix, that is, $V = k' I_p$ with $k' $ a positive constant. \tcr{When $k'$ is very large, the identity component dominates the combined term and weakens the contribution of the sample covariance information.} Similar construction is considered in \cite{jiang2022two} which takes $k' = [np\lambda_{\max}( S_n)]/\varepsilon_n $ for $\varepsilon_n \to 0$ and $\lambda_{\max}(\cdot)$ calculates the largest eigenvalue of a matrix.

The expression of the test statistic in \eqref{eq: bayes factor} requires the joint densities of $X_1$ and $X_2$ under the null and the alternative hypotheses. Under the null, the joint density follows 
\begin{eqnarray}\label{eq: joint density null}
	(2\pi)^{-\frac{np}{2}}|\Sigma|^{-\frac{n}{2}}\exp\left\{-\frac{1}{2}\mbox{tr}\Sigma^{-1}\left[ A_n+n_1(\mu-\bar{ X_1})(\mu-\bar{ X_1})^\top +n_2(\mu-\bar{X_2})(\mu-\bar{X_2})^\top \right]\right\}, \nonumber\\    
\end{eqnarray}
where $A_n = (n-2) S_n$. Further, under the alternative, the joint density follows 
\begin{eqnarray}\label{eq: joint density alternative}
	(2\pi)^{-\frac{np}{2}}|\Sigma|^{-\frac{n}{2}}\exp\left\{-\frac{1}{2}\mbox{tr}\Sigma^{-1}\left[ A_n+n_1(\mu_1-\bar{ X_1})(\mu_1-\bar{X_1})^\top +n_2(\mu_2-\bar{ X_2})(\mu_2-\bar{ X_2})^\top \right]\right\}. \nonumber \\
\end{eqnarray}
By integrating the product of \eqref{eq: joint density alternative} and \eqref{eq: prior alternative} w.r.t. the parameters $\Theta$, the closed-form expression of the numerator of \eqref{eq: bayes factor definition} follows 
\begin{eqnarray}\label{eq: numerator Bayes factor}
	f(X_{1,1}, \ldots, X_{1, n_1}, X_{2, 1}, \ldots, X_{2, n_2} \mid H_{1}) = c_{13} | A_n + V|^{-\frac{1}{2}(m_1+n-2)},   
\end{eqnarray}
where $c_{13}=2^{\frac{(m_1+n-2)p}{2}}\Gamma_p\left(\frac{m_1+n-2}{2}\right)c_{12}$, $c_{12}=(2\pi)^p(n_1n_2)^{-\frac{p}{2}}c_{11}$, $c_{11}=\frac{(2\pi)^{-\frac{np}{2}}|V|^{\frac{m_1}{2}}}{2^{\frac{m_1 p}{2}}\Gamma_p\left(\frac{m_1}{2}\right)}$, and $\Gamma_p(\cdot)$ is a multivariate Gamma function defined in Definition~\ref{def: multivariate Gamma function}.

Similarly, the denominator of \eqref{eq: bayes factor definition} follows
\begin{eqnarray}\label{eq: denominator Bayes factor}
	m(X_{1,1}, \ldots, X_{1, n_1}, X_{2, 1}, \ldots, X_{2, n_2} \mid H_{0}) =c_0| A_n + V+ n_0 D D^\top |^{-\frac{m_0+n-1}{2}},
\end{eqnarray}
where $D = \bar X_1 - \bar X_2$, $n_0 = n_1n_2/n$, and
$c_0=(2\pi)^{-\frac{p(n-1)}{2}}n^{-\frac{p}{2}}| V|^{\frac{m_0}{2}}2^\frac{(n-1)p}{2}\frac{\Gamma_p\left(\frac{m_0+n-1}{2}\right)}{\Gamma_p\left(\frac{m_0}{2}\right)}$.

By taking the ratio, the Bayes factor follows
\begin{eqnarray}\label{eq: bayes factor}
	\lefteqn{\mbox{BF}_{10} =\left(\frac{\pi}{n_0}\right)^{\frac{p}{2}}| V|\frac{\Gamma_p\left(\frac{m_0}{2}\right)}{\Gamma_p\left(\frac{m_1}{2}\right)}\left(\frac{| A_n+  V|}{| A_n + V+n_0 D D^\top|}\right)^{-\frac{m+n}{2}}} &&\\
	&=&\left(\frac{\pi}{n_0}\right)^{\frac{p}{2}}| V|\frac{\Gamma_p\left(\frac{m_0}{2}\right)}{\Gamma_p\left(\frac{m_1}{2}\right)}\left(1+n_0 D^\top  (A_n+V)^{-1} D\right)^{\frac{m+n}{2}}.
\end{eqnarray}

Notice in \eqref{eq: bayes factor}, when $V$ is noninformative and does not carry information of the samples, $\mbox{BF}_{10}$ is monotonically increasing with respect to $n_{0} {D}^\top(A_n + k'I_p)^{-1} D = n_{0} {D}^\top(S_n + kI_p)^{-1} D$, \tcr{where $k = \frac{1}{n-2} k'$. We therefore use $n_{0} {D}^\top(S_n + kI_p)^{-1} D$ as the key quantity for discriminating $H_0$ and $H_1$. It is worth noting that the scalar degrees of freedom $m_0$ and $m_1$ in the inverse-Wishart prior distribution do not appear in the above expression and hence do not affect the resulting test statistic.}
\tcr{Since $S_n$ is unstable in high dimensions, directly using $(S_n + kI_p)^{-1}$ may still be inaccurate. Moreover, Lemma~\ref{lem: converge of matrix} with proof in Appendix~\ref{proof of lemma 1} identifies an over-regularized regime for the choice of $k$. Specifically, if $k = O(p^2)$, then
\[
(S_n + kI_p)^{-1} - \frac{1}{k}I_p \to 0.
\]
That is, $(S_n + kI_p)^{-1}$ becomes asymptotically equivalent to a scaled identity matrix. Therefore, choosing $k$ of order $O(p^2)$ would over-shrink the covariance structure and dilute the variance information contained in $S_n$. The lemma is thus used to identify an undesirable regime, rather than to motivate such a choice of $k$.}

To retain stable variance information while avoiding unreliable off-diagonal estimation, we instead focus on $\mathrm{diag}(S_n)$ and define
\[
\Lambda_n = (\mathrm{diag}(S_n) + kI_p)^{-1}.
\]
\tcr{The subsequent analysis is established for any $k = o(p^2)$, and in Section~\ref{sec: numerical} we take $k = 1$, which is equivalent to specifying an inverse-Wishart prior with scale matrix $V = k'I_p = (n-2)I_p$. This choice yields a prior with large dispersion, so that the prior is weakly informative.
}

\section{Test statistic} \label{sec: Test statistic}
We focus on the population version of $\Lambda_n$ denoted by $\Lambda = (\mbox{diag}(\Sigma) + k I_p)^{-1}$ with components $(1 / ([\Sigma]_{jj} +k))$ and $\Sigma$ is the common population covariance matrix. \tcr{We study the expectation and variance of the quadratic form that results from replacing the sample covariance with the population covariance in the Bayes factor in \eqref{eq: expectation} and \eqref{eq: variance}. This intermediate step justifies the centering and scaling of the proposed statistic and simplifies the subsequent asymptotic analysis.} Under $H_0$ in \eqref{eq: null hypothesis}, the expectation and variance of $n_{0}{D}^\top \Lambda{D}$ are determined by the Gaussian random vector $D$ as follows
\begin{eqnarray}
	&&E(n_{0}{D}^\top\Lambda {D})=\mbox{tr}(\Lambda{\Sigma})=\sum_{i=1}^{p} \frac{[\Sigma]_{ii}}{[\Sigma]_{ii}+k}, \label{eq: expectation}\\
	&&\mbox{Var}(n_{0}{D}^\top \Lambda {D})=2\mbox{tr}(\Lambda \Sigma)^2. \label{eq: variance}
\end{eqnarray}

To simplify the notation, we denote a matrix $U =\Lambda{\Sigma}$. Intuitively, by the central limit theorem, the following asymptotic normality holds 
\begin{eqnarray}\label{eq: CLT}
	n_{0}{D}^\top \Lambda D - \mbox{tr}(U) \stackrel{d}{\to} N(0, 2\mbox{tr}(U^2)),
\end{eqnarray}
where the $j$th diagonal element of $U =\Lambda{\Sigma}$ is $[\Sigma]_{jj}/([\Sigma]_{jj} + k) < 1$. 

Our proposed test statistic is constructed based on \eqref{eq: CLT} by replacing the unknown terms $\Lambda$ and $\Sigma$ with their sample versions $\Lambda_n$ and $S_n$, that is 
\begin{eqnarray}\label{eq: test stats}
	T_{{\rm BF, 1}}=\frac{n_{0}{D}^\top \Lambda_n {D}- \mbox{tr}(U_n)}{\sqrt{2
			\widehat{\mbox{tr}(U^2)}}},
\end{eqnarray}
where $U_n = \Lambda_n S_n$ and the consistent estimator 
$\widehat{\mbox{tr}(U^2)}$ is given in Proposition~\ref{prop: consistent estimators}. We will show in Proposition~\ref{prop: consistent estimators} that the mean and variance estimators $\mbox{tr}(U_n)$ and $\widehat{\mbox{tr}(U^2)}$ are ratio consistent. Further, we will establish in Theorem~\ref{thm: asymptotic normality} the asymptotic normality of 
\[\frac{n_{0}{D}^\top \Lambda_n {D}- \mbox{tr}(U_n)}{\sqrt{2 \mbox{tr}(U^2)}}.\] 
By ratio consistency of $\mbox{tr}(U^2)$ and $\widehat{\mbox{tr}(U^2)}$, the asymptotic normality holds for the test statistic \eqref{eq: test stats}. A small sample corrected version of \eqref{eq: test stats} will be proposed in Section \ref{small sample correction}.

\subsection{Ratio consistent estimators}
Proposition~\ref{prop: consistent estimators} provides the ratio consistent estimators for $\mbox{tr}({U})$ and $\mbox{tr}(U^2)$, and subsequently investigate the relationship between $n_{0} {D}^\top \Lambda {D}$ and $n_{0}{D}^\top \Lambda_n {D}$. 

\begin{proposition}\label{prop: consistent estimators}
	Under Assumption~\ref{cond: bounded eigen value}, the ratio consistent estimators of $\mbox{tr}(U)$ and $\mbox{tr}( U^2)$ are, respectively, given by
	\begin{enumerate}
		\item $\widehat{\mbox{tr}({U})}=\mbox{tr}({U_n})$;
		\item $\widehat{\mbox{tr}(U^2)}=\mbox{tr}{(U_n)^2}-\frac{1}{n-2}[\mbox{tr}({U_n})]^2$,
	\end{enumerate}
\end{proposition}

The proof of Proposition~\ref{prop: consistent estimators} is provided in Section~\ref{ssec: proof of proposition 1}. The resulting ratio-consistent estimators can be directly used in the test statistic in \eqref{eq: test stats}. \tcr{Notice that 
\begin{equation}\label{eq: test stat split}
\frac{n_{0} D^\top \Lambda_n D-\mbox{tr}(U_n)}{\sqrt{2\mbox{tr}(U^2)}}
=
\frac{n_{0}D^\top \Lambda_n D-\mbox{tr}(\Lambda_n\Sigma)}{\sqrt{2\mbox{tr}(U^2)}}
+
\frac{\mbox{tr}(\Lambda_n\Sigma)-\mbox{tr}(U_n)}{\sqrt{2\mbox{tr}(U^2)}}.
\end{equation}
To establish the asymptotic normality of the left-hand side of \eqref{eq: test stat split}, we first show that the difference between the first term on the right-hand side and its population counterpart is asymptotically negligible, see Lemma~\ref{lem: lambda replace} with proof in Appendix~\ref{ssec: proof of lemma lambda replace}.}  

\tcr{
\begin{lemma}\label{lem: lambda replace}
Under Assumptions~\ref{cond: bounded eigen value}, \ref{cond: negligible}, and \ref{cond: random mat},
\begin{eqnarray}\label{eq: approxi error}
\frac{n_{0}D^\top(\Lambda_n-\Lambda)D-\mbox{tr}\{(\Lambda_n-\Lambda)\Sigma\}}
{\sqrt{2\mbox{tr}(U^2)}} \stackrel{P}{\to} 0
\end{eqnarray}
under $H_0$. Furthermore, if
\[
n_0\delta^\top(\Lambda_n-\Lambda)\delta=o_P\bigl(\sqrt{\mbox{tr}(U^2)}\bigr),
\]
where $\delta=\mu_1-\mu_2$, then \eqref{eq: approxi error} also holds under $H_1$.
\end{lemma}
}

With the guarantee from Lemma~\ref{lem: lambda replace}, we can approximate the first term on the right-hand-side of \eqref{eq: test stat split} by $\frac{n_{0}D^\top \Lambda D-\mbox{tr}(\Lambda\Sigma)}{\sqrt{2\mbox{tr}(U^2)}}$ and conclude 
\begin{equation}\label{eq: test stat split1}
\frac{n_{0} D^\top \Lambda_n D-\mbox{tr}(U_n)}{\sqrt{2\mbox{tr}(U^2)}}
-
\frac{n_{0}D^\top \Lambda D-\mbox{tr}(\Lambda\Sigma)}{\sqrt{2\mbox{tr}(U^2)}}
-
\frac{\mbox{tr}(\Lambda_n\Sigma)-\mbox{tr}(U_n)}{\sqrt{2\mbox{tr}(U^2)}}
\stackrel{P}{\to}0.
\end{equation}

To proceed to the main result in Theorem~\ref{thm: asymptotic normality}, we first show in Lemma~\ref{prop: remainder converge to zero} that the third term in \eqref{eq: test stat split1} is negligible.

\begin{lemma}\label{prop: remainder converge to zero}
	Under Assumptions~\ref{cond: bounded eigen value}, \ref{cond: Q}, it holds that
	$$\frac{\mbox{tr}(\Lambda_n\Sigma)-\mbox{tr}( U_n)}{\sqrt{2\mbox{tr}(U^2)}}\to 0.$$
\end{lemma}

The proof of Lemma~\ref{prop: remainder converge to zero} is included in Section~\ref{ssec: proof of prop 2}. It follows from \eqref{eq: test stat split} and Lemma~\ref{prop: remainder converge to zero}, that
\begin{equation}\label{eq: test stat split2}
	\frac{n_{0} {D}^\top \Lambda_n {D}-\mbox{tr}(U_n)}{\sqrt{2\mbox{tr}(U^2)}}-\frac{n_{0}{D}^\top \Lambda {D}-\mbox{tr}(U)}{\sqrt{2\mbox{tr}( U^2)}} = O_P(n^{-1/2}),
\end{equation}
which converges to 0 as $n\to \infty$. Thus, to obtain the asymptotic normality of the first term in (\ref{eq: test stat split2}), it suffices to show the asymptotic normality of the second term. The results are summarized in Theorem~\ref{thm: asymptotic normality}.
\begin{theorem}\label{thm: asymptotic normality}
	Under Assumptions~\ref{A1} and \ref{cond: negligible}, the following asymptotic normality of $n_0 D^\top \Lambda_n D$ holds
	\begin{enumerate}
		\item  Under $H_0$,
		\begin{equation}
			\frac{n_{0}{D}^\top \Lambda_n {D}-\mbox{tr}(U_n)}{\sqrt{2\mbox{tr}(U^2)}}\stackrel{d}{\to} N(0,1).
		\end{equation}
		\item Under $H_1$,
		\begin{equation}
			\frac{n_{0} D^\top \Lambda_n  D-\mbox{tr}(U_n)-n_{0} \delta^\top \Lambda \delta}{\sqrt{2\mbox{tr}( U^2)}} \stackrel{d}{\to} N(0,1).
		\end{equation}
	\end{enumerate}	
\end{theorem}

The proof of Theorem~\ref{thm: asymptotic normality} is included in Section~\ref{ssec: proof of theorem 1}. The power of the proposed test is presented in Corollary~\ref{Cor: power of the test} as a function of $\delta = \mu_1 - \mu_2$. 

\begin{corollary}\label{Cor: power of the test}
	Let $g(\delta)$ be the power of the test $T_{BF}$. Under Assumption~\ref{cond: random mat}, 
	$$g(\delta)- \Phi\left(-u_{1-\alpha}+\frac{n_{0}\delta^\top \Lambda\delta}{\sqrt{2{\mbox{tr}(U^2)}}}\right)\to0,$$
	where $\Phi(x)$ is the cumulative distribution function of the standard normal distribution, and $\Phi(u_{1-\alpha}) = 1-\alpha$.
\end{corollary}

\subsection{Small sample correction}\label{small sample correction}
It can be seen from Theorem~\ref{thm: asymptotic normality} that the test statistic \eqref{eq: test stats} is asymptotically normal under the null hypothesis $H_0$ and the local alternative $H_1$ for large $n$. However, the remainder terms of the $\sqrt{n}$-consistent estimators in \eqref{eq: test stats} are not negligible when the sample size $n$ is small. We address the performance of the test statistic in small samples by adding bias-correction coefficients to account for the coordinates of the sample mean vectors. We denote the bias correction coefficients as $R_{1} = {\rm{diag}}({r}_{11}, {r}_{21}, \cdots, {r}_{p1})$, $R_{2} = {\rm{diag}}({r}_{12}, {r}_{22}, \cdots, {r}_{p2})$, $R_{3} = {\rm{diag}}({r}_{13}, {r}_{23}, \cdots, {r}_{p3})$, where for $j=1,2,\ldots, p$,
\begin{equation*}
r_{j1}=E\left(\frac{[\Sigma]_{jj}+k}{[S_n]_{jj}+k}\right), \quad
r_{j2}=\frac{[\Sigma]_{jj}+k}{[\Sigma]_{jj}}E\left(\frac{[S_n]_{jj}}{[S_n]_{jj}+k}\right),
\end{equation*}
\[
r_{j3}=\frac{n-2}{n-3}\left(\frac{[\Sigma]_{jj}}{[\Sigma]_{jj}+k}\right)^2 \left(E\left(\frac{[S_n]_{jj}}{[S_n]_{jj}+k}\right)^2\right)^{-1}.
\]
With $R_1, R_2, R_3$, the test statistic for small samples is as follows 
\begin{equation}\label{eq: test stats small sample}
T_{\rm{BF}, 2}=\frac{n_{0}{D}^\top \Lambda_nR_{1}^{-1} {D}-\mbox{tr}(R_{2}^{-1}U_n)}{\sqrt{2 \widetilde{\mathrm{tr}(U^2)}}},
\end{equation}
where $\widetilde{\mathrm{tr}(U^2)}$ is a ratio consistent estimator of $\mathrm{tr}(U^2)$ is given by
\begin{equation}
\widetilde{\mathrm{tr}(U^2)}=\mathrm{tr}({U_n^*}^2)-\frac{1}{n-2}[\mathrm{tr}(U_n^*)]^2,
\end{equation}
\tcr{and $U_n^*$ denotes the bias-corrected version of $U_n$, obtained by replacing the diagonal terms in $U_n$ with their corrected counterparts induced by $R_3$.}

The correction coefficients $R_1$ and $R_2$ are obtained by matching $n_0 D^\top \Lambda_n D$, $\mbox{tr}(U_n)$ with their population quantity $n_{0}{D}^\top \Lambda D$ and $tr(U)$, respectively. \tcr{All Monte Carlo results reported below are based on this bias-corrected version unless stated otherwise.} And the correction for $\mbox{tr}(U^2)$ is based on using the terms including $\left(\frac{[S_n]_{jj}}{[S_n]_{jj}+k}\right)^2$ in the bias corrected estimator of $\mbox{tr}(\Sigma^2)$ in \cite{bai1996effect} following
\begin{equation}
B_n^2=\frac{(n-2)^2}{n(n-3)}\left[\mathrm{tr}(S_n^2)-\frac{1}{n-2}(\mathrm{tr}S_n)^2\right].
\end{equation}

\begin{remark}
In practice, the expectations can be calculated numerically. Taking for example $r_{j1} = E\left(\frac{[\Sigma]_{jj}+k}{[S_n]_{jj}+k}\right)$, we rewrite it as $E\left(\frac{n-2+a_j}{Y_j+a_j}\right)$, where $Y_j=\frac{(n-2)[S_n]_{jj}}{[\Sigma]_{jj}}\sim\chi^2_{n-2}$ and $a_j=\frac{k(n-2)}{[\Sigma]_{jj}}$. Given $a_j$, the expectation is easy to get by using numerical computation. Notice that there is an unknown parameter $\frac{1}{[\Sigma]_{jj}}$  in $a_j$, we replace it by its unbiased estimator $\frac{h}{[S_n]_{jj}}$, where $h=\frac{n-4}{n-2}$. 
\end{remark}

\section{Numerical performance}\label{sec: numerical}
We assess the effectiveness of the Bayes factor-based test across different scenarios. 
The significance level is set to $\alpha = 0.05$ in all simulations. The data $X_{1,1}, \ldots, X_{1,n_1}$ and $X_{2,1}, \ldots, X_{2,n_2}$ are generated from multivariate normal distributions $N_p(\mu_1, \Sigma)$ and $N_p(\mu_2, \Sigma)$. We consider four configurations for the covariance matrix $\Sigma$, where the first two are from \cite{zoh2018powerful} and the last from \cite{srivastava2016raptt}.

\begin{enumerate}
	\item $\Sigma_1 = I_p$. 
	\item $\Sigma_2$ has elements $\sigma_{ij} = 0.4^{|i-j|}$.
	\item $\Sigma_3$ has diagonal elements $\lambda_j = 12/j$ for $j = 1, 2, \ldots, 12$, and
	\[
	\Sigma_{3, (j,j')} = \begin{cases}
		\frac{12}{j} I\{j \leq 12\} + I\{j > 12\}, & j = j' \\
		0.3, & |j - j'| = 1 \\
		0.1^k, & |j - j'| > 1
	\end{cases}
	\]
	\item $\Sigma_4 = \mathrm{diag}(\lambda_1, \ldots, \lambda_p)$, where $\lambda_j = 20/j$ for $j = 1, 2, \ldots, 20$ and $\lambda_j = 1$ for $j = 21, \ldots, p$.
\end{enumerate}

Without loss of generality, for the null hypothesis, we let $\mu_1 = \mu_2 = 0$. For the alternative, we set $\mu_1 = 0$, and consider two possible alternatives as follows \cite{zoh2018powerful}

\begin{enumerate}
	\item {Alternative 1}: We set the original $\mu_2 \sim N({1}, {\mathrm{I}}_p)$, select $p_0 \times p$ randomly chosen elements to be zero, \tcr{where $p_0$ is the sparsity level, defined as the proportion of null components in the mean vector and a larger $p_0$ corresponds to a sparser alternative, while a smaller $p_0$ corresponds to a denser one.} We then scale $\mu_2$ such that $\delta^{\mathrm{T}} \Sigma^{-1} \delta = 2$.
	\item {Alternative 2}: We first set $\mu_2$ and $p_0$ the same as in Alternative 1, and then scale $\mu_2$ such that $\delta^{\mathrm{T}} \delta = 0.1 \sqrt{\mathrm{tr}(\Sigma^2)}$.
\end{enumerate}

We compared the size and power of the proposed test with several competitive tests in the existing literature: \tcr{BS in \cite{bai1996effect}, which is based on the squared Euclidean distance between the sample means under equal covariance matrices; CQ in \cite{chen2010two}, which modifies BS to handle unequal covariance matrices; SD in \cite{srivastava2008test}, which replaces the full covariance matrix by its diagonal to avoid singularity; PB in \cite{jiang2022two}, which is constructed from the posterior Bayes factor; and RM in \cite{zoh2018powerful}, which is based on random projection.} As discussed above, Lemma~\ref{lem: converge of matrix}, we take $k = 1$, equivalently $k' = (n-2)$, for the proposed BF test and $k' = n(\log n) p\lambda_{\max} ((n-2)S_n)$ for the PB test by \cite{jiang2022two} throughout the simulation. 
\tcr{The sparsity level $p_0$ is set to $0.5, 0.6, 0.7, 0.8,$ and $0.9$, where $p_0 = 0.5$ corresponds to the least sparse case with 50\% nulls, and $p_0 = 0.9$ corresponds to the most sparse case with 90\% nulls.}

The simulation results averaged over $R = 5000$ replications and the nominal level $\alpha = 0.05$. 
\tcr{The tests are evaluated through the empirical rejection function, defined as
\begin{eqnarray*}
    \hat{\beta}(\alpha)= \frac{1}{R} \sum_{r = 1}^R I\{ T^{(r)} \ge u_{1-\alpha} \},
\end{eqnarray*}
where $R$ is the number of Monte Carlo replications, $T^{(r)}$ denotes the test statistics at the $r$'th replication, and $u_{1-\alpha} = \Phi^{-1}(1 - \alpha)$ with $\Phi^{-1}(\cdot)$ the standard normal quantile function. This quantity corresponds to the empirical size when the data are generated under the null hypothesis $H_0:\mu_1=\mu_2=0$, and to the empirical power when the data are generated under the alternative hypothesis.}

of the tests are all close to the nominal level $\alpha = 0.05$ and are summarized in Table~\ref{table:test_sizes}. The change of the empirical power of the tests against the sparsity level $p_0$ is presented in Figure~\ref{fig4}, and the tables are included in Section~\ref{section: simulation tables} in the Appendix. Figure~\ref{fig4} shows that the tests have similar powers under $\Sigma_1$ and $\Sigma_2$. Upon closer inspection, the BF, PB, BS, and CQ tests perform relatively well in both scenarios. Under $\Sigma_3$ and $\Sigma_4$, however, the powers of the SD and BF tests are much higher than those of the PB, BS, and CQ tests, with the BF test generally performing better than the SD test. The RM test exhibits lower power than the other tests under $\Sigma_1$ and $\Sigma_2$, but under $\Sigma_3$ and $\Sigma_4$, its power is generally between that of the BF and BS tests. 

\begin{figure}[hp!]
	\centering
	\begin{minipage}{0.45\textwidth}\includegraphics[width=\textwidth,height=4.6cm]{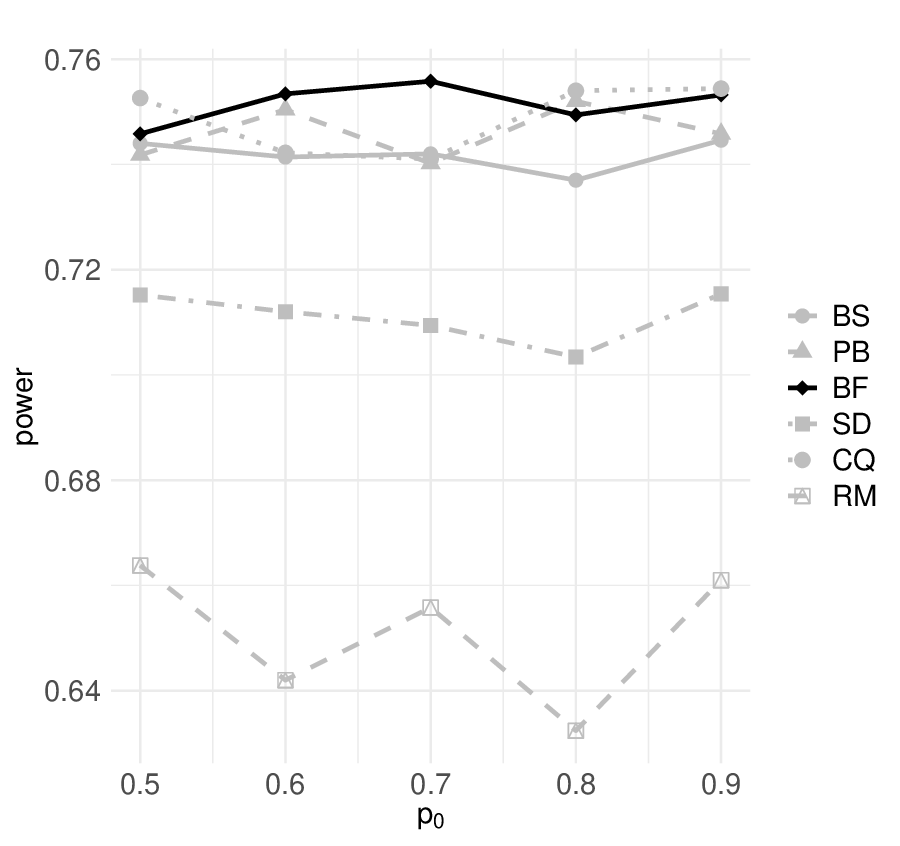}
	\end{minipage} \hfill
	\begin{minipage}{0.45\textwidth}
		\includegraphics[width=\textwidth,height=4.6cm]{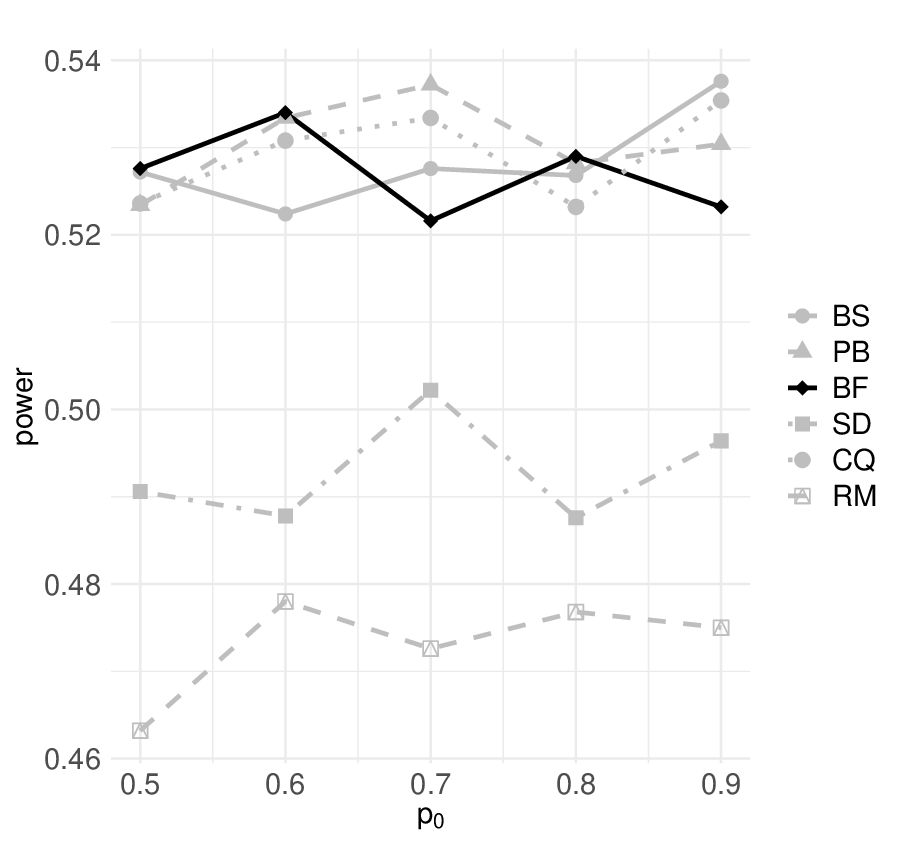}
	\end{minipage}
	\begin{minipage}{0.45\textwidth}
		\includegraphics[width=\textwidth,height=4.6cm]{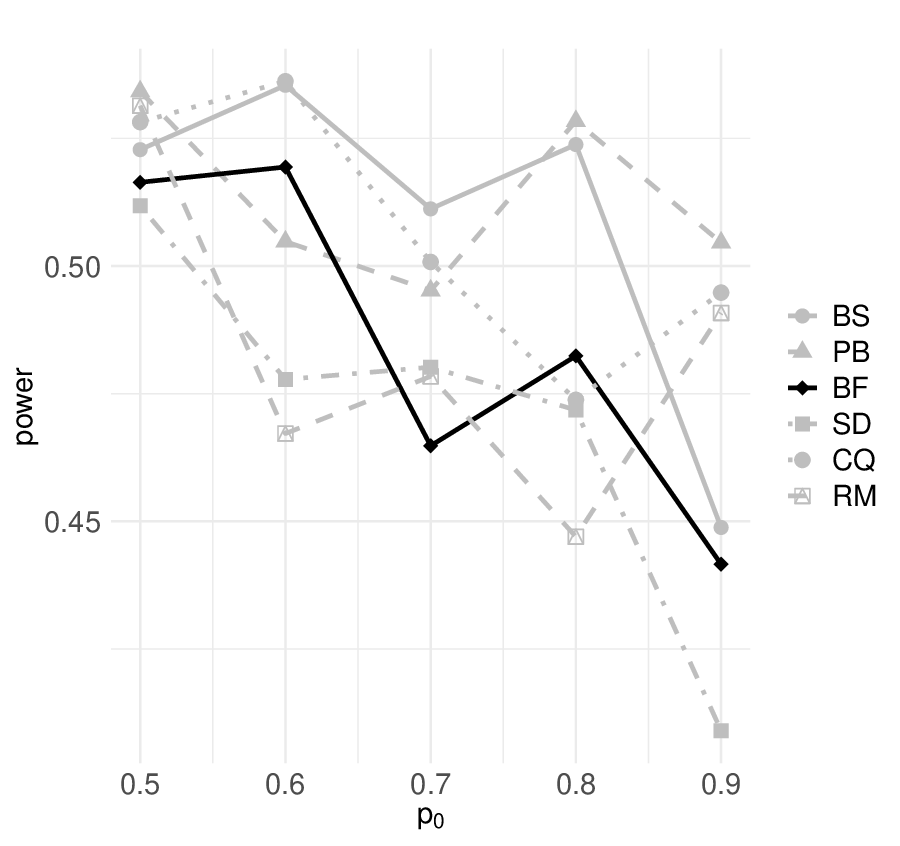}
	\end{minipage} \hfill
	\begin{minipage}{0.45\textwidth}
		\includegraphics[width=\textwidth,height=4.6cm]{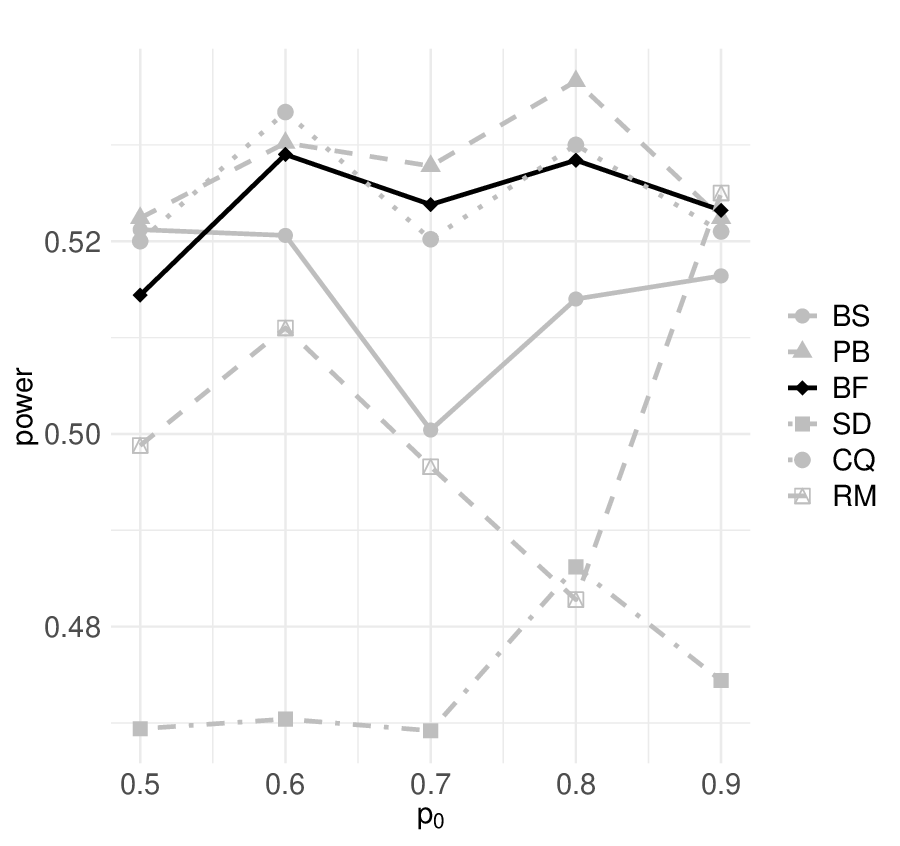}
	\end{minipage}
	\begin{minipage}{0.45\textwidth}
		\includegraphics[width=\textwidth,height=4.6cm]{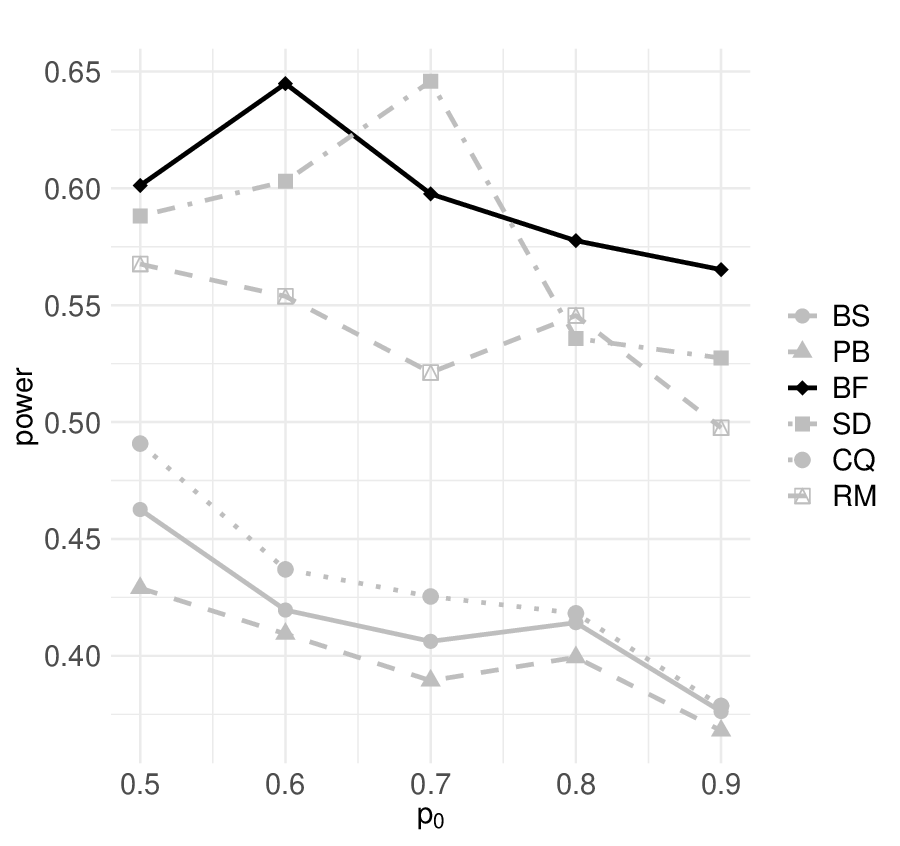}
	\end{minipage} \hfill
	\begin{minipage}{0.45\textwidth}
		\includegraphics[width=\textwidth,height=4.6cm]{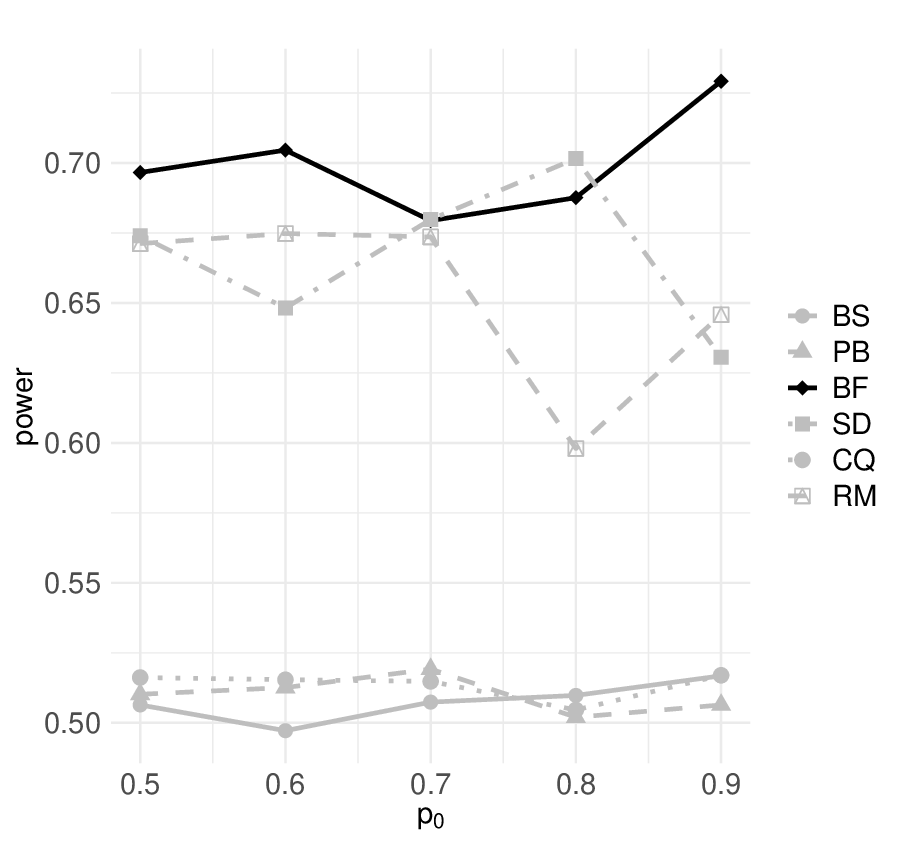}
	\end{minipage}
	\begin{minipage}{0.45\textwidth}
		\includegraphics[width=\textwidth,height=4.6cm]{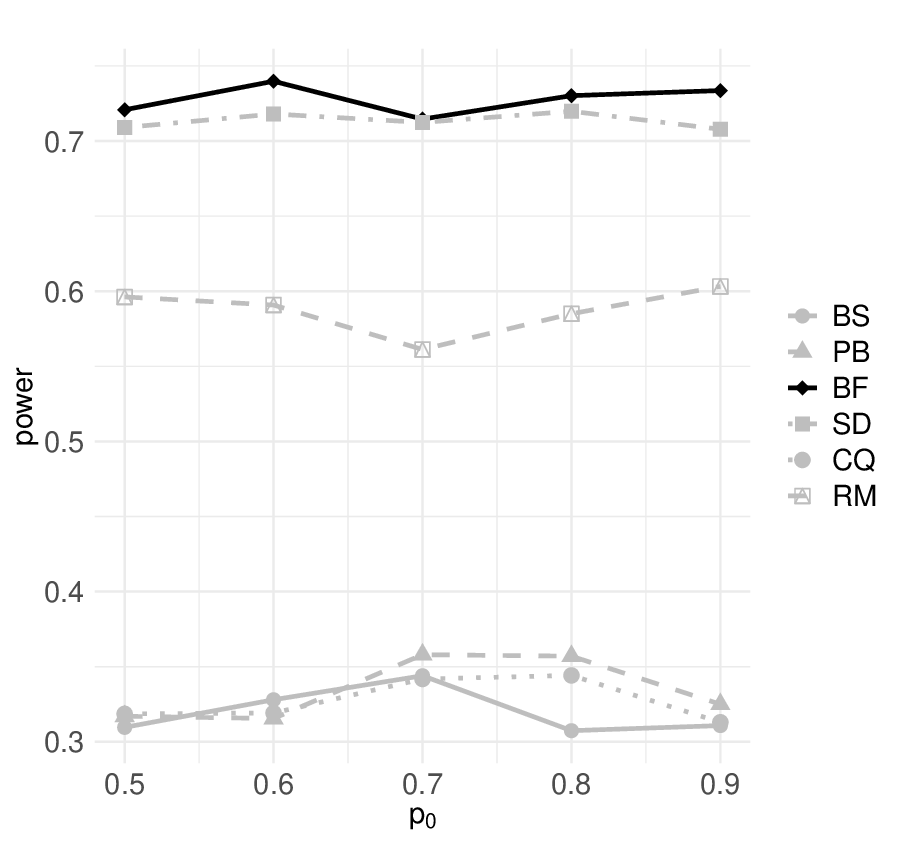}
	\end{minipage} \hfill
	\begin{minipage}{0.45\textwidth}
		\includegraphics[width=\textwidth,height=4.6cm]{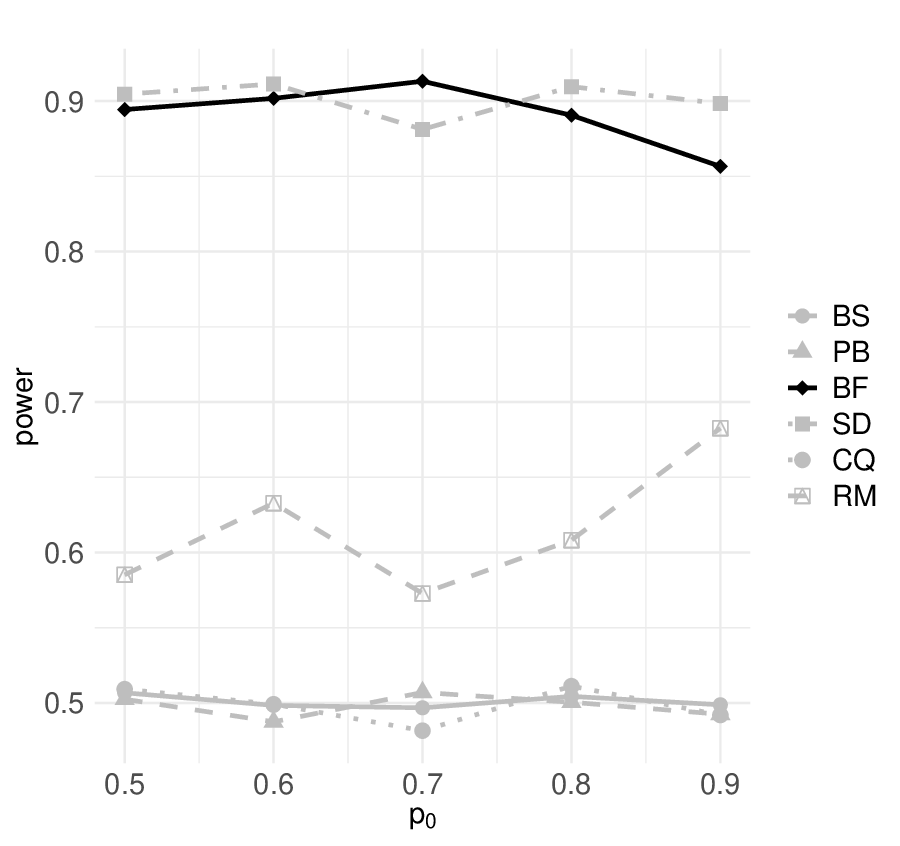}
	\end{minipage}
	\caption{Power plots (y-axis) for the tests under varying sparsity levels (x-axis) $p_0 = 0.5, 0.6, 0.7, 0.8, 0.9$, where $p_0$ denotes the proportion of null components in the mean vector. Each row corresponds to a different covariance matrix, $\Sigma = \Sigma_1, \Sigma_2, \Sigma_3, \Sigma_4$. The left panel presents results under Alternative 1, while the right panel shows results under Alternative 2. The proposed test is highlighted in black.}
	\label{fig4}  
\end{figure}

\subsection{Small sample performance}

To demonstrate the stable performance of the proposed BF test in small-sample settings, we consider the case where $n_1 = n_2 = 5$. We set ${\mu}_1 = {0}$ and investigate the power of the test by setting the first 25\% of the components of ${\mu}_2$ to 0.5, the last 25\% to -0.5, and the remaining to 0. The main finding is that the proposed BF test controls its size well, whereas the competing tests produce size values significantly higher than the nominal level, sometimes exceeding 0.08. 

From Table~\ref{table:small_sample_performance}, the powers of the BF and SD tests are very close, with the BF test showing slightly higher power. However, the BF test yields smaller size values than the SD test. When $n_1 = n_2 = 5$, the BF test controls its size well, whereas the other tests all have inflated empirical sizes. 
In most cases, the BS, PB, and CQ tests exhibit similar sizes and powers. The RM test has lower power than the other five tests under ${\Sigma}_1$ and ${\Sigma}_2$, whereas under ${\Sigma}_3$ and ${\Sigma}_4$, its power lies between the BF and BS tests. 

We also tested the performance of the tests in small samples in various sparsity levels where $p_0 = 0.8, 0.9$. Test results are similar to those in Table~\ref{table:small_sample_performance}. Our proposed test has a stable and competitive performance among all.

\begin{table}[hbtp!]
	\centering
	\begin{tabular}{c c c c c c| c c c c }
		\toprule
		& & \multicolumn{4}{c}{$p=160$} & \multicolumn{4}{c}{$p=200$} \\
		\cmidrule(lr){3-6} \cmidrule(lr){7-10}
		$n_1=n_2=5$& & $\Sigma_1$ & $\Sigma_2$ & $\Sigma_3$ & $\Sigma_4$ & $\Sigma_1$ & $\Sigma_2$ & $\Sigma_3$ & $\Sigma_4$ \\
		\midrule
		\multirow{2}{*}{BF} & Size & 0.0528 & 0.0562 & 0.0538 & 0.0568 & 0.0552 & 0.0560 & 0.0566 & 0.0568 \\
		& Power & 0.7584 & 0.6258 & 0.6430 & 0.6494 & 0.8138 & 0.6974 & 0.7162 & 0.7404 \\
		\midrule
		\multirow{2}{*}{SD} & Size & 0.0794 & 0.0856 & 0.0780 & 0.0812 & 0.0854 & 0.0778 & 0.0774 & 0.0776 \\
		& Power & 0.6960 & 0.6058 & 0.6120 & 0.6168 & 0.7560 & 0.6570 & 0.6766 & 0.7002 \\
		\midrule
		\multirow{2}{*}{PB} & Size & 0.0682 & 0.0662 & 0.0760 & 0.0860 & 0.0630 & 0.0710 & 0.0814 & 0.0884 \\
		& Power & 0.7862 & 0.6670 & 0.5232 & 0.3708 & 0.8512 & 0.7332 & 0.6040 & 0.4482 \\
		\hline
		\multirow{2}{*}{CQ}	&size&0.0652&0.0674&0.0758&0.0836&0.0624&0.0662&0.0736&0.0806\\
		&power&0.7836&0.6496&0.4990&0.3320&0.8522&0.7200&0.5864&0.4442\\
		\hline
		\multirow{2}{*}{BS}	&size&0.0620&0.0684&0.0836&0.0878&0.0582&0.0612&0.0762&0.0846\\
		&power&0.7860&0.6682&0.5048&0.3508&0.8480&0.7232&0.5976&0.4470\\
		\hline
		\multirow{2}{*}{RM}	&size&0.0596&0.0768&0.0818&0.0840&0.0632&0.0762&0.0812&0.0878\\
		&power&0.6418&0.5862&0.5232&0.4832&0.7166&0.6634&0.6102&0.5694\\
		\bottomrule
	\end{tabular}
	\caption{The power and size of the tests in small samples where the sample size $n_1 = n_2 = 5$.}
	\label{table:small_sample_performance}
\end{table}

\subsection{Gaussianity misspecification}
We also want to test the proposed test when the two samples do not follow multivariate Gaussian distributions. \tcr{These experiments are intended solely to assess empirical robustness; providing a theoretical justification is beyond the scope of this paper.} We consider the following 
\begin{equation}
	{X}_{1i} = {\mu_1} + {\Sigma}^{\frac{1}{2}} {Z}_i, \quad i=1,2,\dots,n_1;
	\quad {X}_{2 i'} = {\mu_2} + {\Sigma}^{\frac{1}{2}} {Z}_{i'}, \quad i'=1,2,\dots,n_2,
\end{equation}
where the components of $Z_i$ and $Z_{i'}$ are independent and follows $(\chi_2^2 - 2)/2$. 
In addition to $\Sigma_4$ which describes independent coordinates of $X_{1i}$s and $X_{2i'}$s, we consider ${\Sigma}_5 = {D}^{\frac{1}{2}} {\Omega} {D}^{\frac{1}{2}}$, where the elements of ${D}$ are independently drawn from $U(1,3)$, and ${\Omega}$ has components $\Omega_{ij} = 0.6^{|i-j|}$.

To obtain the power of the tests under misspecification, we set $\mu_1 = 0$. And for ${\mu}_2$, we set the first 10\% to 0.4, the last 10\% to $-0.4$, and the remaining elements to zero.

\begin{figure}[htbp!]
	\centering
	\begin{minipage}{0.48\textwidth}
		\centering
		\includegraphics[width=\textwidth]{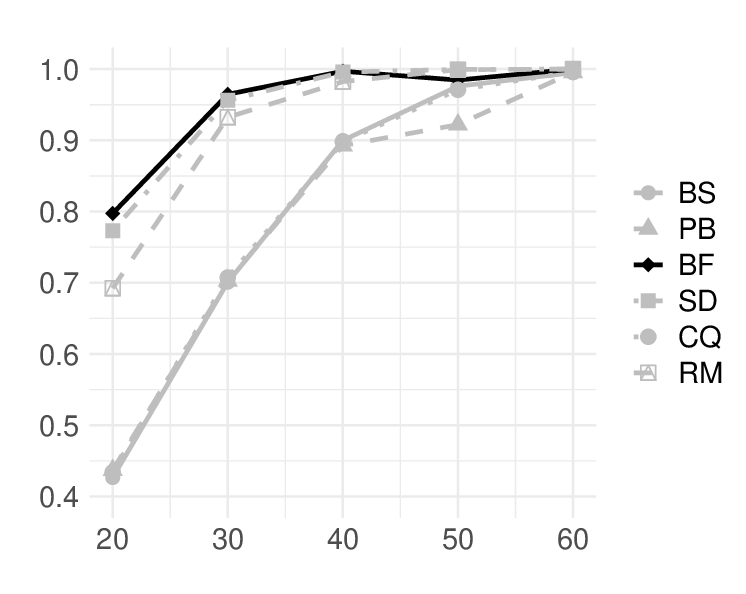}
	\end{minipage} \hfill
	\begin{minipage}{0.48\textwidth}
		\centering
		\includegraphics[width=\textwidth]{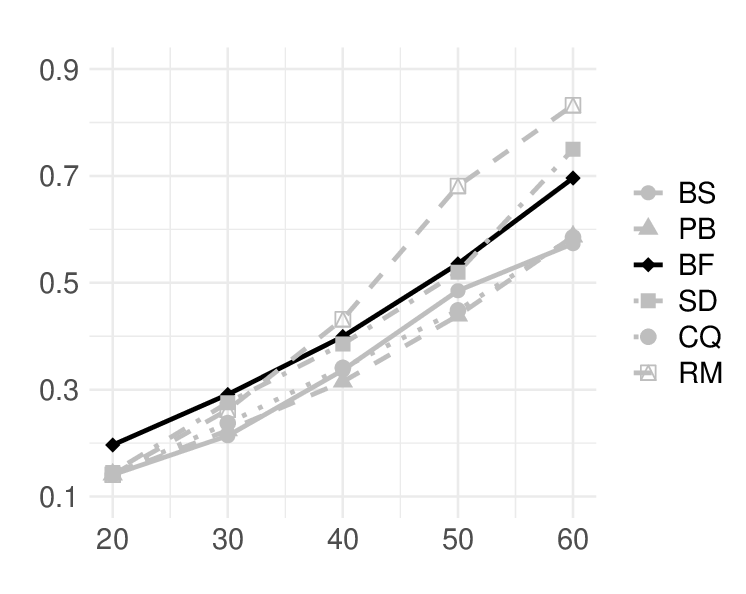}
	\end{minipage}
	\caption{Power (y-axis) of the tests for sample sizes $n_1 = n_2 = 20, 30, 40, 50, 60$ (x-axis), with $p = 200$. The left panel corresponds to $\Sigma_4$, and the right panel corresponds to $\Sigma_5$. The proposed test is highlighted in black.}
	\label{fig: distribution misspecification}
\end{figure}

The size and power of the tests are recorded in Tables~\ref{table: misspecified Gaussian sigma 4} and \ref{table: misspecified Gaussian sigma 5} in Section~\ref{section: simulation tables} in the Appendix. The sizes of all tests are near the nominal level under distribution misspecification. The powers of the tests can be found in Figure~\ref{fig: distribution misspecification}, we observe that the proposed BF test has superior performance than the competing tests for $\Sigma_4$ and small sample for $\Sigma_5$,\; while the RM test has better performance when $n_1=n_2=40,50,60$ under $\Sigma_5$.

\begin{figure}[hbtp!]
	\centering
	\begin{minipage}{0.48\textwidth}
		\includegraphics[width=\textwidth,height=5.2cm]{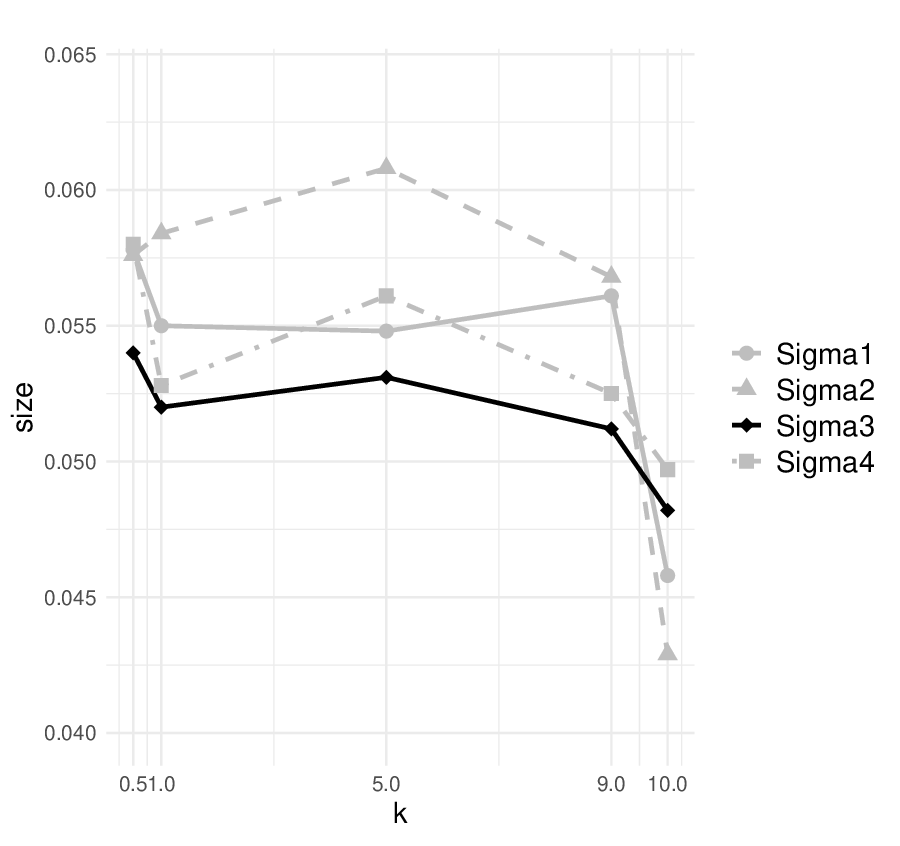}
	\end{minipage}
	\begin{minipage}{0.48\textwidth}
		\includegraphics[width=\textwidth,height=5.2cm]{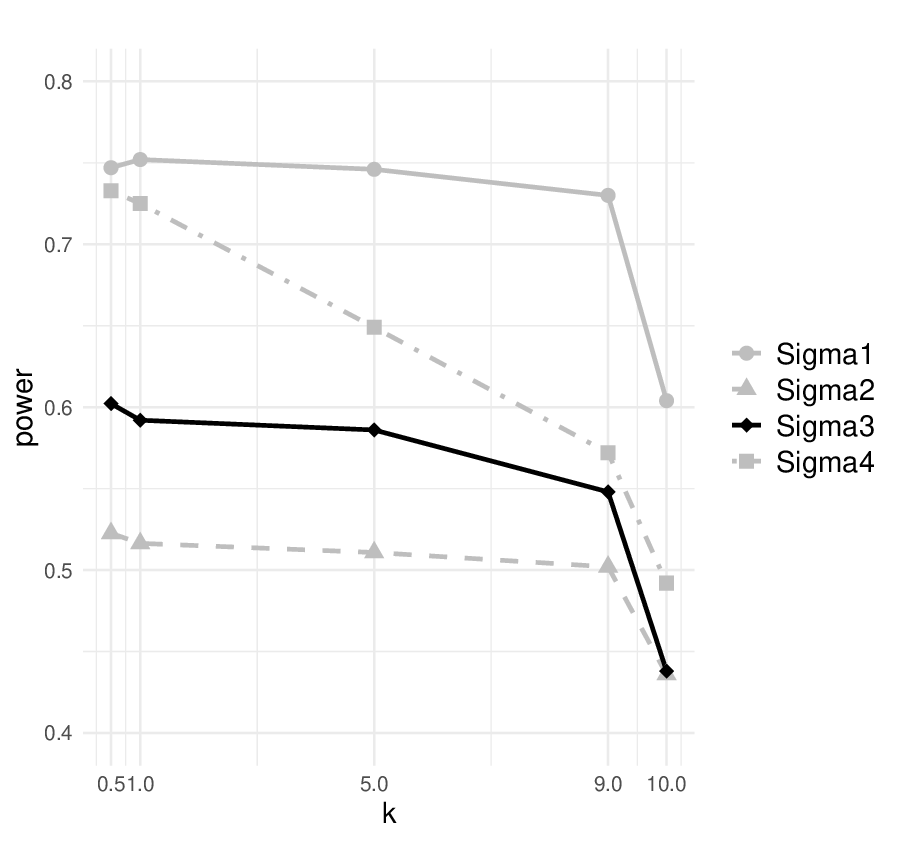}
	\end{minipage} \hfill
	\begin{minipage}{0.48\textwidth}
		\includegraphics[width=\textwidth,height=5.2cm]{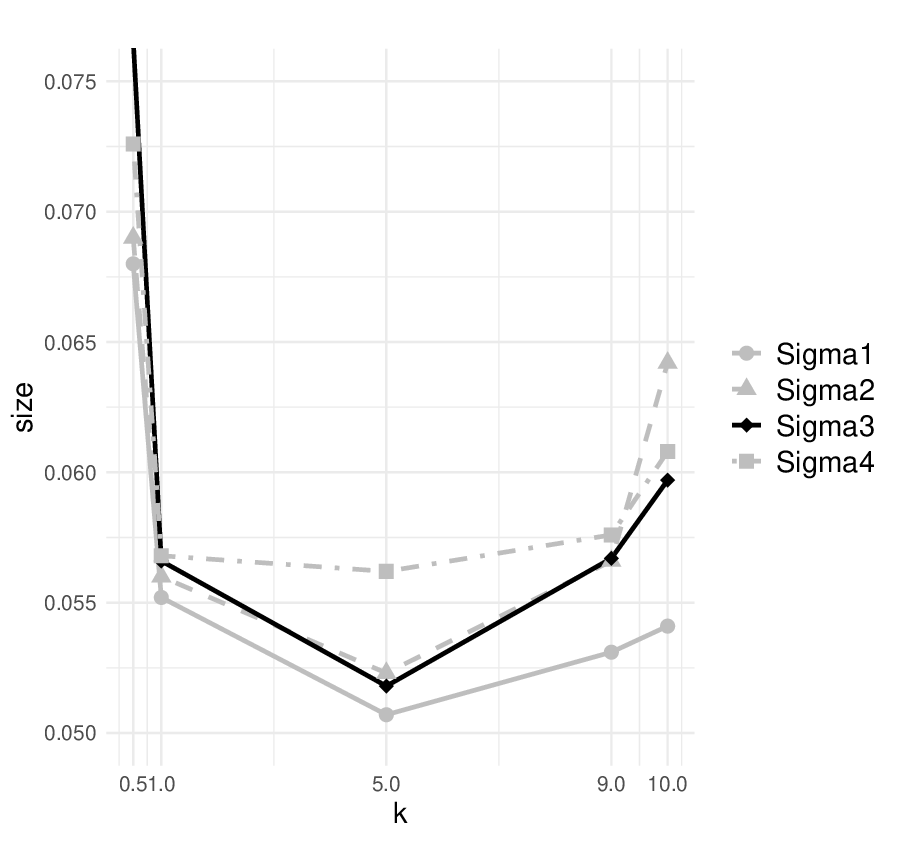}
	\end{minipage}
	\begin{minipage}{0.48\textwidth}
    \includegraphics[width=\textwidth,height=5.2cm]{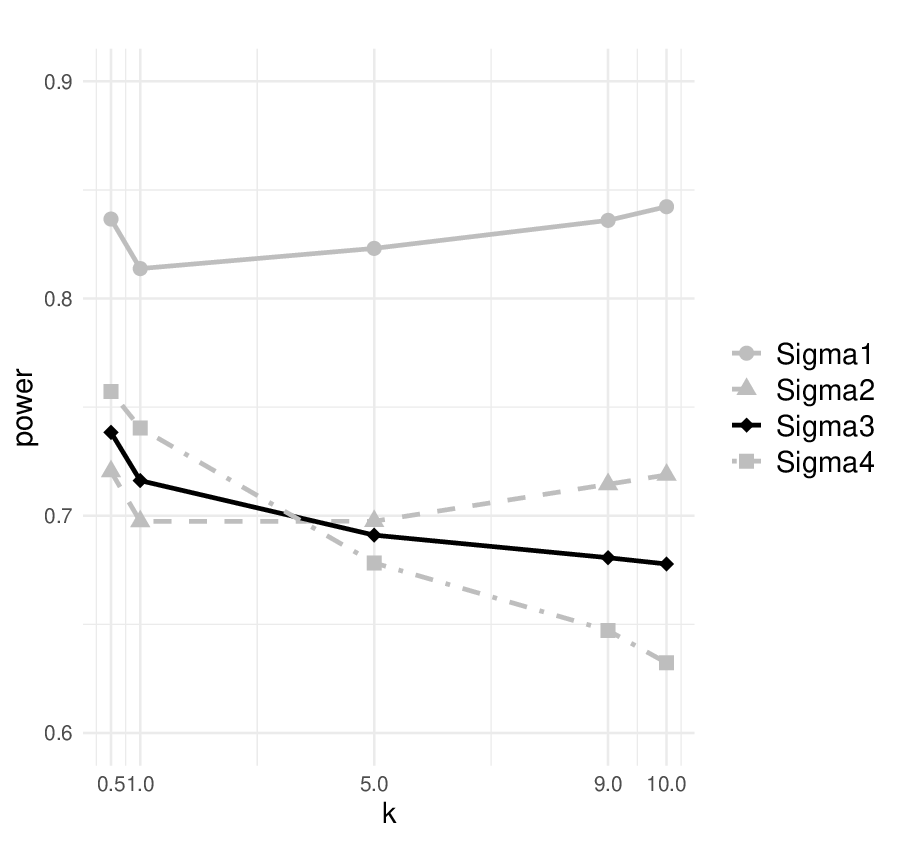}
	\end{minipage} \hfill
	\begin{minipage}{0.48\textwidth}
		\includegraphics[width=\textwidth,height=5.2cm]{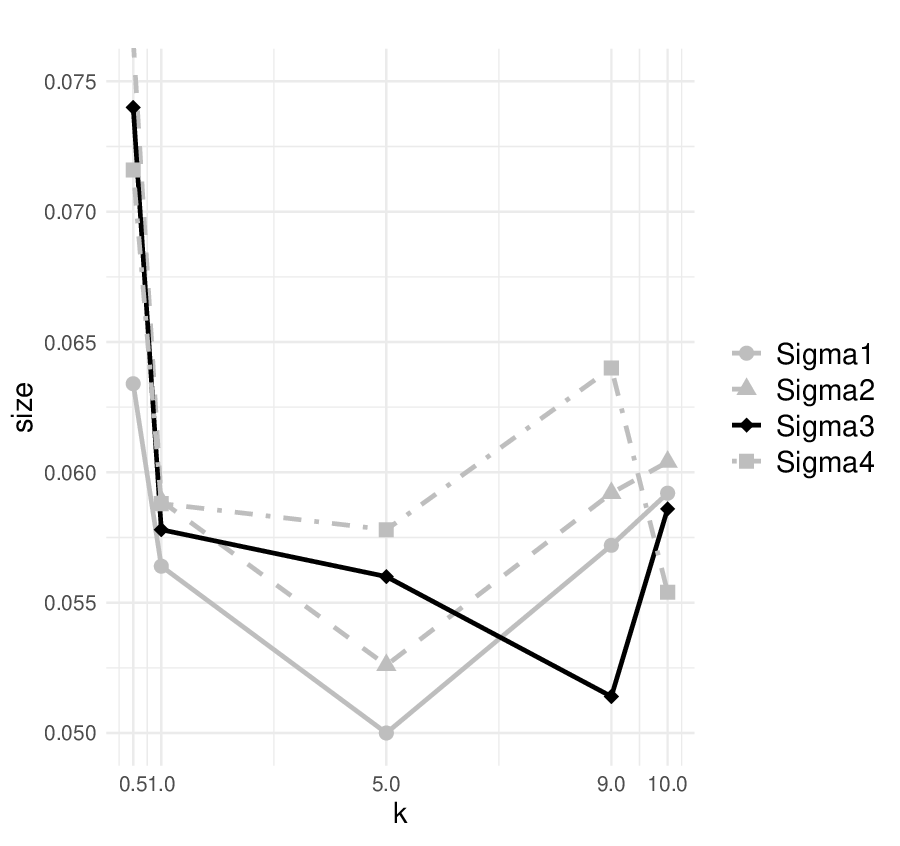}
	\end{minipage}
	\begin{minipage}{0.48\textwidth}
		\includegraphics[width=\textwidth,height=5.2cm]{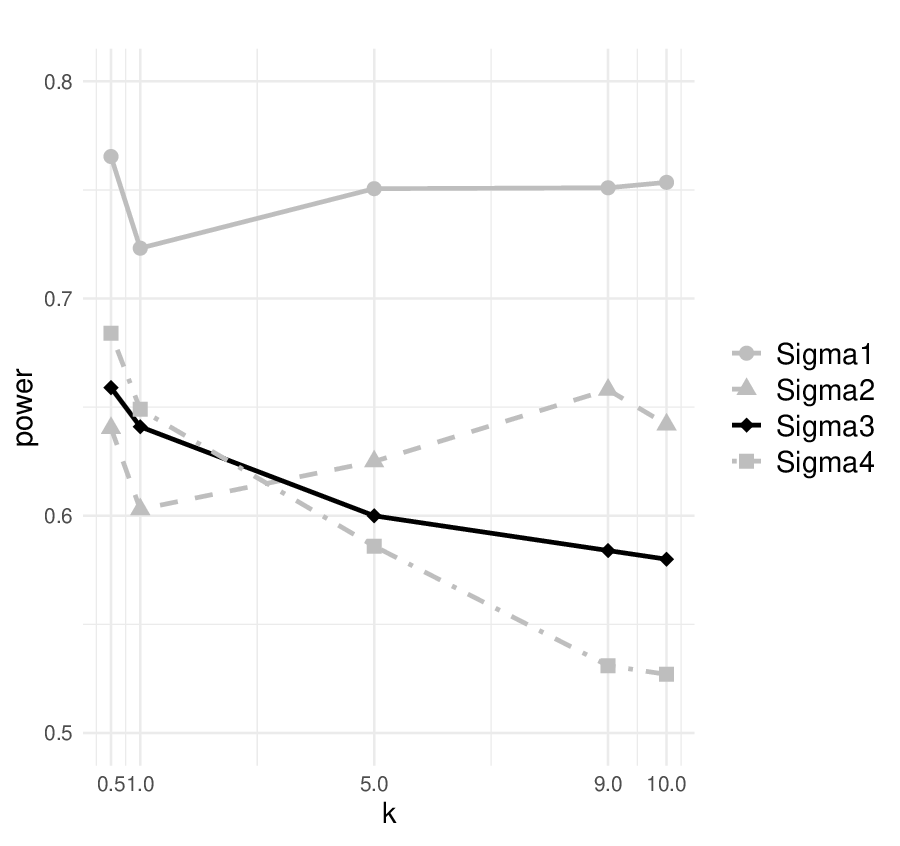}
	\end{minipage} 
	\caption{Empirical size (left) and power (right) of the proposed test for $k \in \{0.5,1,5,9,10\}$ under three representative settings. The first row corresponds to $(p,n_1,n_2)=(200,50,50)$ , Alternative 1, and $p_0=0.5$. The second and third rows correspond to $(p,n_1,n_2)=(200,5,5)$ and $(160,5,5)$, respectively. 
    }
	\label{fig: sensitivity}  
\end{figure}

\tcr{
\subsection{The sensitivity of $k$}
To investigate the sensitivity of the proposed procedure to the choice of $k$, we consider $k \in \{0.5,1,5,9,10\}$. Figure~\ref{fig: sensitivity} reports the empirical size and power under three representative settings. In the first setting, $p=200$, $n_1=n_2=50$, Alternative 1 is used, and $p_0=0.5$. The two panels in the first row display the corresponding empirical size and power. The second and third rows report the results for $(p,n_1,n_2)=(200,5,5)$ and $(160,5,5)$, respectively.}

\tcr{
When $p=200$ and $n_1=n_2=50$, the empirical power decreases as $k$ increases, and the empirical size also exhibits an overall downward trend. When $p=200$ and $n_1=n_2=5$, the empirical size decreases sharply for small values of $k$ and then increases gradually, whereas the empirical power shows different trends across variance settings. Similar behavior is also observed when $p=160$ and $n_1=n_2=5$. }

\tcr{
The observed sensitivity to $k$ is not unexpected, since it directly affects the key element of the test statistic
$\Lambda_n=(\mbox{diag}(S_n)+kI_p)^{-1}$.
As a result, different choices of $k$ lead to different levels of shrinkage and may influence the empirical size and power of the test. 
}

\section{Small Round Blue Cell Tumors dataset} \label{sec: real data analysis}

We test the proposed Bayes factor-based test on the small round blue cell tumors (SRBCTs) dataset \footnote{https://file.biolab.si/biolab/supp/bi-cancer/projections/info/SRBCT.html}. SRBCTs consist of four different small tumors: neuroblastoma (NB), Ewings family of tumors (EWS), rhabdomyosarcoma (RMS), and non-Hodgkin lymphoma (BL), with sample sizes of 18, 29, 25, and 11, respectively. Each sample has $p = 2308$ gene expressions. To test whether gene expression differs between pairs, we randomly select two from the four, resulting in 6 pairs in total, and report the test results. The abbreviations are as follows: we use ``NR" to represent the pair NB and RMS, ``ER" to represent the pair EWS and RMS, and the other abbreviations follow a similar pattern. \tcr{The main purpose of this application is to illustrate that the proposed statistic is computationally feasible and can be applied to real data in a setting where $p \gg n$. Since datasets such as SRBCT typically have relatively small sample sizes and the degree of sparsity in gene expression differences across tumor types is not known a priori, this example offers a relevant setting for practical comparison.}

We compare the test statistics. Since all tests reject the null hypothesis with $p$-values less than $10^{-5}$, we report $\log p$-values in parentheses for readability. \tcr{Table~\ref{tab:SRBCT} further shows that the proposed procedure can be implemented on a high-dimensional gene expression dataset and yields conclusions comparable to those of existing methods. Further, the proposed test produces more extreme test statistics for several pairs.}

\begin{table}[ht!]
	\centering
	\begin{tabular}{c c c c c c}
		\toprule
		&BF&PB&SD&BS&CQ \\
		\midrule
		NR&12.83 (-75.41)&13.27 (-91.57)& 6.84 (-26.28)&13.27 (-91.57)&\cellcolor{lightgray}13.50 (-94.69)\\
		\hline
		ER&11.74 (-72.34)&\cellcolor{lightgray}14.19 (-104.19) &5.57 (-18.16) &\cellcolor{lightgray}14.19 (-104.19) &14.16 (-103.77) \\
		\hline
		BN& \cellcolor{lightgray}14.50 (-108.74)  & 11.55 (-70.06) & 7.81 (-33.49)& 11.55 (-70.06) & 11.35 (-67.81)\\
		\hline
		BE&\cellcolor{lightgray}14.80 (-113.07)&11.91 (-74.29)&8.17 (-36.44) &11.91 (-74.29) &12.01 (-75.59)\\
		\hline
		BR&\cellcolor{lightgray}18.71 (-178.87) &14.91 (-144.79) &11.13 (-65.33) &14.91 (-144.79) &15.11 (-117.82) \\
		\hline
		NE&14.13 (-103.39)\cellcolor{lightgray} & 11.88 (-73.95)& 8.13 (-36.06)& 11.88 (-73.95) & 12.13 (-76.96)\\
		\bottomrule
	\end{tabular}
	\caption{Test statistics and $\log$ $p$-values for SRBCTs.}
	\label{tab:SRBCT}
\end{table}



\section{Conclusion} \label{sec:conc}
We proposed a test statistic based on the Bayes factor and applied it to test the differences between group means in high-dimensional settings. We established the asymptotic normality of the test and derived its asymptotic power. Through simulation studies, we demonstrated the effectiveness and superiority of the proposed test over existing methods, suggesting that it performs robustly across different scenarios.

\tcr{It is worth emphasizing that the current derivation and all formal proofs rely on Gaussianity. Although the numerical experiments indicate a degree of robustness under misspecification, those experiments do not provide theoretical guarantees beyond the Gaussian setting. Extending the method to weaker assumptions, such as spherical or more general distributions, would require new marginal-likelihood calculations and separate asymptotic arguments, and remains an important direction for future work.}

\tcr{Another important direction is computational efficiency. Although the final test statistic depends only on diagonal quantities, the implementation still requires repeated construction of pooled sample covariance, diagonal regularization, and bias calibration across Monte Carlo settings. In our simulations, the running time at dimension 200 is roughly three to four times that at dimension 100 under otherwise identical settings, so the computational burden is not negligible in genuinely high-dimensional problems. Developing more efficient algorithms for large-scale datasets, and adapting the approach to multiple groups or online testing with additional covariates, would further extend the applicability of the test.}

\clearpage
\begin{appendices}

\section{Assumptions and useful results}\label{secA1}
\subsection{Assumptions}
\begin{enumerate}
	\item \label{A1} $\lim_{p\to\infty}\mbox{tr}(U^4)/(\mbox{tr} U^2)^2=0$.
	\item \label{cond: bounded eigen value} $0<m_1=\min([\Sigma]_{jj})\leqslant \max([\Sigma]_{jj})=m_2<\infty$\;uniformly in $p$.
	\item \label{cond: negligible} $n_0 \delta^\top \Lambda\Sigma \Lambda \delta=o(\mathrm{tr}( U^2))$,\;as $n,p \to\infty$, where $\delta=\mu_1- \mu_2$.
	\item \label{cond: random mat} $p/n\to c_1\in(0,\infty)$,\;$n_1/n\to c_2\in(0,1)$ as $n,p \to\infty$.
	\item \label{cond: Q} $0<\lim_{p\to\infty}\mbox{tr}( Q^i)/p<\infty,\;i=1,2,3,4$, where $Q= \Sigma^{\frac{1}{2}}\Lambda \Sigma^{\frac{1}{2}}$.
\end{enumerate}

\begin{remark}
    \tcr{Assumption~\ref{cond: bounded eigen value} means that the marginal variances of the coordinates remain finite and bounded away from zero; Assumption~\ref{cond: random mat} means that the dimension and sample size grow at comparable rates, so the problem is high-dimensional but not overwhelmingly more high-dimensional than the available sample size; and Assumptions~\ref{A1}, \ref{cond: negligible}, and \ref{cond: Q} are technical regularity conditions used to control higher-order terms in the proofs.} Further, Assumptions~\ref{A1}, \ref{cond: negligible} are similar to (A2) and assumption (B) in \cite{srivastava2013two}; \ref{cond: bounded eigen value} is the (A1) in \cite{srivastava2013two}, while \ref{cond: Q} is similar to (2.3) in \cite{srivastava2008test}. 
\end{remark}

\subsection{Useful results}
\begin{definition}\label{def: multivariate Gamma function}
	A multivariate Gamma function $\Gamma_p(\cdot)$ is defined as
	\begin{equation*}
		\Gamma_p(x)=\pi^{p(p-1)/4}\prod_{j=1}^p\Gamma[x+(1-j)/2],
	\end{equation*}
	where $\Gamma(x) = \int_0^\infty t^{x-1}e^{-t} dt$ is the Gamma function.
\end{definition}

\begin{lemma}\label{lem: converge of matrix}
	Under Assumptions~\ref{cond: bounded eigen value}, \ref{cond: random mat}, when $k = O(p^2)$,
	\begin{eqnarray}
		(S_n + k I_p)^{-1}-\frac{1}{k} I_p \to 0.
	\end{eqnarray}
\end{lemma}

\begin{lemma}\label{lem: convergence} (\cite{shiryaev2016probability})
	For a mutually independent sequence of random variables $\xi_1,\cdots,\xi_n$, $E \xi_i=a_i$, $\mbox{Var}\xi_i=b_i^2$, $B_n^2=\sum\limits_{i=1}^n b_i^2$, if there exists a positive number $\delta$ such that
	$$\frac{1}{B_n^{2+\delta}}\sum\limits_{i=1}^{n}{E}|\xi_i-a_i|^{2+\delta}\rightarrow0, n\rightarrow\infty,$$
	then
	$$P\left\{\frac{1}{B_n}\sum\limits_{i=1}^n(\xi_i-a_i)<x\right\}\rightarrow\Phi(x).$$
\end{lemma}

\begin{lemma}\label{lem: rewrite covariance matrix}(\cite{anderson1958introduction})
	If $X_1,\cdots,X_n$ constitute an independent sample from $N_p(\mu,\Sigma)$, define $\bar X=\sum_{i=1}^n X_i$ and sample covariance matrix $\widehat{\Sigma}=\frac{1}{n}\sum_{i=1}^{n}(X_i-\bar{ X})(X_i-\bar{ X})^\top$. Then, $n\widehat{\Sigma}$ is distributed as $\sum_{i=1}^{n-1} Z_i Z_i^\top$, where $Z_i\sim N_p(0,\Sigma)$ are independent Gaussian vectors.
\end{lemma}

\begin{lemma}\label{lem: moments of standard Gaussian}
	The $w$th moments of a standard normal random variable $X$ are as follows 
	\begin{equation*}
		{E}(X^w)=
		\begin{cases}
			\frac{\Gamma(\frac{w+1}{2})2^\frac{w+1}{2}}{\sqrt{2\pi}} &  w =2t,\\
			0 \quad\quad\quad\quad  & w=2t+1,\\
		\end{cases}
	\end{equation*}
	where $t=0,1,2,\cdots.$
\end{lemma}

\begin{lemma}\label{lem: convergence in prob}
	Let $X_j$, $j \ge 1$ be a sequence of random variables. If $X_n \stackrel{L^r}{\to }{ X}$ for some $r > 0$, then $X_n \stackrel{P}{\to} X$.
\end{lemma}

	\section{Proofs}\label{secB1}
\subsection{Proof of Lemma~\ref{lem: converge of matrix}}\label{proof of lemma 1}
To prove
\begin{equation*}
	(S_n + k I_p)^{-1}-\frac{1}{k} I_p \to 0,
\end{equation*}
we only need to prove 
\begin{equation*}
	\frac{S_n}{k}\to  0.
\end{equation*} 
Recall the pooled covariance matrix
\begin{equation*}
	S_n=\frac{1}{n-2}\left\{(n_1-1) S_1+(n_2-1) S_2\right\},
\end{equation*}
with the elements
\begin{equation}\label{eq: sij}
	[S_n]_{ij}=\frac{1}{n-2}\big\{(n_1-1)[S_{n, 1}]_{ij}+(n_2-1)[S_{n_2}]_{ij}\big\},
\end{equation}
where $S_1$ and $S_2$ are the covariance matrices of the two samples with elements $[S_{n_1}]_{ij}$ and $[S_{n_2}]_{ij}$.
\begin{eqnarray}\label{eq: s_n ij}
	[S_n]_{ij}^2 =\frac{1}{(n-2)^2}\{(n_1-1)^2 [S_{n, 1}]_{ij}^2+(n_2-1)^2 [S_{n,2}]_{ij}^2 && \nonumber\\
	\qquad \qquad +2(n_1-1)(n_2-1)[S_{n,1 }]_{ij} [S_{n,2}]_{ij}\}
\end{eqnarray}
\begin{eqnarray}\label{eq: sn product}
	[S_n]_{ii}[S_n]_{jj}&=\frac{1}{(n-2)^2}\{(n_1-1)[S_{n, 1}]_{ii}+(n_2-1) [S_{n, 2}]_{ii}][(n_1-1)[S_{n, 1}]_{jj} \nonumber\\
	& +(n_2-1)[S_{n, 2}]_{jj}\} \nonumber\\
	&=\frac{1}{(n-2)^2}\left\{(n_1-1)^2[S_{n, 1}]_{ii}[S_{n, 1}]_{jj}+(n_2-1)^2[S_{n, 2}]_{ii}[S_{n, 2}]_{jj}\right\} \nonumber\\
	&+\frac{1}{(n-2)^2}\left\{(n_1-1)(n_2-1)([S_{n, 1}]_{ii}[S_{n, 2}]_{jj}+[S_{n, 2}]_{ii}[S_{n, 1}]_{jj})\right\}.
\end{eqnarray}

Now, we only need to show $\|{S_n}\|_F/k \to 0$ when $k = O(p^2)$.

\begin{equation}
	\frac{\Vert{S_n}\Vert_F^2}{k^2}=\sum_{i,j=1}^p\frac{[S_{n}]_{ij}^2}{k^2}=\sum_{i,j=1}^p\frac{s_{ij}^2}{p^4}.
\end{equation}
By the property of variance-covariance between two random variables, the following two inequalities hold 
\begin{equation}\label{eq: variance covariance inequality 1}
	[S_{n, 1}]_{ij}\le \sqrt{ [S_{n, 1}]_{ii} [S_{n, 1}]_{jj}},    
\end{equation}
\begin{equation}\label{eq: variance covariance inequality 2}
	[S_{n, 2}]_{ij}\le \sqrt{ [S_{n,2}]_{ii}  [S_{n, 2}]_{jj}},      
\end{equation}
where the right-hand sides are the square root of the products of variance terms. 
By \eqref{eq: variance covariance inequality 1} and \eqref{eq: variance 
	covariance inequality 2}, the following inequality holds
\begin{eqnarray}\label{eq: arithmetic and geometric means}
	2[S_{n,1}]_{ij} [S_{n, 2}]_{ij} & \le 2\sqrt{[S_{n, 1}]_{ii} [S_{n, 1}]_{jj}}[S_{n,2}]_{ii}  [S_{n, 2}]_{jj}& \nonumber\\
	& \le [S_{n, 1}]_{ii} [S_{n, 2}]_{jj} + [S_{n,2}]_{ii}[S_{n, 1}]_{jj},
\end{eqnarray}
where the first equality is obtained by multiplying \eqref{eq: variance covariance inequality 1} and \eqref{eq: variance covariance inequality 2}, and the second inequality holds by the inequality of arithmetic and geometric means.

By using \eqref{eq: arithmetic and geometric means}, \eqref{eq: s_n ij}, \eqref{eq: sn product}, it is easy to show that 
$$[S_{n}]_{ij}^2\leq [S_{n}]_{ii}[S_{n}]_{jj},$$ 
which suggests
\begin{equation}\label{eq: frobenius norm}
	\left(\frac{\Vert{S_n}\Vert_F}{k}\right)^2 = \sum_{i,j=1}^p\frac{[S_{n}]_{ij}^2}{p^4}\leq\frac{(\sum_{i=1}^p [S_{n}]_{ii})^2}{p^4} = \left(\frac{(\sum_{i=1}^p [S_{n}]_{ii})}{p^2}\right)^2.
\end{equation}
We know that $\frac{(n_1-1)[S_{n, 1}]_{ii}}{[\Sigma]_{ii}}\sim\chi^2(n_1-1)$ and $\frac{(n_2-1)[S_{n, 2}]_{ii}}{[\Sigma]_{ii}}\sim\chi^2(n_2-1)$, it follows that 
$$(n-2)[S_n]_{ii}/[\Sigma]_{ii}\sim\chi^2(n-2).$$

Since $[S_n]_{ii}$'s are unbiased estimators of $[\Sigma]_{ii}$, we obtain
$${E}\left(\sum_{i=1}^p[S_n]_{ii}\right)=\mbox{tr}(\Sigma) = O (p),$$
and
$$\mbox{Var}\left(\sum_{i=1}^p[S_n]_{ii}\right)\leq\left(\sum_{i=1}^p\sqrt{\mbox{Var}([S_n]_{ii})}\right)^2=\frac{2}{n-2}(\mathrm{tr}(\Sigma))^2 =\frac{2}{n - 2} O(p^2) = O(p).$$
Under \ref{cond: bounded eigen value}, the right side of \eqref{eq: frobenius norm} tends to $0$ as $p \to \infty$ since both the mean and variance of the random variable tend to zero. Hence, we conclude that \[\frac{\|S_n\|_F}{k} \to 0 .\]

\subsection{Proof of Proposition~\ref{prop: consistent estimators}}\label{ssec: proof of proposition 1}
\begin{enumerate}
	\item 
	It holds that
	\[\mbox{tr}(U)=\mbox{tr}(I_p-  k\Lambda),\]
	\[\mbox{tr}(U_n)=\mbox{tr}(I_p- k\Lambda_n),\]
	where the diagonal matrices $\Lambda = (\mbox{diag}(\Sigma) + k I_p)^{-1}$ and $\Lambda_n = (\mbox{diag}(S_n) + k I_p)^{-1}$.
	Thus, we only need to prove 
	$$\frac{{\mbox{tr}(\Lambda)}}{\mbox{tr}(\Lambda_n)} \stackrel{P}{\to} 1.$$
	Following \cite{srivastava2009test} using Talyor expansion to approximate $[S_n]_{ii}$, the $i$th diagonal element of $\Lambda$ can be approximated as follows 
	\begin{equation}\label{eq: taylor expansion}
		\frac{1}{[S_n]_{ii}+k}=\frac{1}{[\Sigma]_{ii}+k}-\frac{[S_n]_{ii}-[\Sigma]_{ii}}{([\Sigma]_{ii}+k)^2}+\frac{([S_n]_{ii}-[\Sigma]_{ii})^2}{([\Sigma]_{ii}+k)^3}+O_P\left(n^{-1}\right),
	\end{equation}

	Since $\frac{(n-2)[S_n]_{ii}}{[\Sigma]_{ii}}\sim\chi^2_{n-2}$,
	\begin{eqnarray}\label{eq: expectations}
		& E [S_n]_{ii} = [\Sigma]_{ii}, \nonumber \\ 
		& E([S_n]_{ii}-[\Sigma]_{ii})^2 = {2[\Sigma]_{ii}^2}/{(n-2)},\nonumber\\
		&\mbox{Var} \left(([S_n]_{ii}-[\Sigma]_{ii})^2\right)={8(n+4)}[\Sigma]_{ii}^4 / {(n-2)^3},\nonumber\\
		& \mbox{Cov}\left([S_n]_{ii}-[\Sigma]_{ii},([S_n]_{ii}-[\Sigma]_{ii})^2\right)={8[\Sigma]_{ii}^3}/{(n-2)^2}.
	\end{eqnarray}
	By combining \eqref{eq: expectations} and \eqref{eq: taylor expansion}, it holds that 
	\begin{eqnarray*}
		&&{E}\left(\frac{1}{[S_n]_{ii}+k}\right)=\frac{1}{[\Sigma]_{ii}+k}+ O\left(n^{-1}\right)  \\
		&&\mbox{Var}\left(\frac{1}{[S_n]_{ii} + k}\right)=O\left(n^{-1}\right),
	\end{eqnarray*}
	which suggests that \[\frac{1}{[S_n]_{ii}+k}=\frac{1}{[\Sigma]_{ii}+k}+O_p\left(n^{-1/2}\right).\]

	From Chebyshev's inequality,
	$$\frac{({[S_n]_{ii}+k})^{-1}}{([\Sigma]_{ii}+k)^{-1}} - 1  = O_P\left(n^{-1/2}\right), i = 1, \ldots, p.$$
	
	Then, we conclude that
	$$\Lambda_n=\left(1+O_P\left(n^{-1/2}\right)\right)\Lambda,$$
	$$\mbox{tr}(\Lambda_n)=\left(1+O_p\left(n^{-1/2}\right)\right)\mbox{tr}(\Lambda).$$	
	\item As $R_3\to I_p$ when $n\to\infty$,\;so,\;we only need to prove it when $R_3$ doesn't exist.\;To prove this, we follow \cite[Proof of Eq.~(16)]{jiang2022two} and first rewrite $S_n$ by Lemma~\ref{lem: rewrite covariance matrix} as follows 
	\begin{equation}\label{eq: Sn rewrite}
		S_n=\frac{1}{n-2}{\sum_{i=1}^{n-2} \Sigma^{{1}/{2}} Z_i  Z_i^\top \Sigma^{{1}/{2}}},
	\end{equation}
	where $ Z_i \sim N_p(0, I_p)$ are independently distributed multivariate standard Gaussian random vectors. By \eqref{eq: Sn rewrite}, the estimator of $\mbox{tr}(U^2)$ can be written as
	\begin{eqnarray*}
		\lefteqn{\mbox{tr}(U_n^2)-\frac{1}{n-2}(\mbox{tr} U_n)^2}&& \\
		&=&\left\{\frac{1}{(n-2)^2}\mbox{tr}\left[\Lambda\sum_{i=1}^{n-2}\Sigma^{\frac{1}{2}} Z_i Z_i^\top \Sigma^{\frac{1}{2}}\right]^2-\frac{1}{(n-2)^3}\left[\sum_{i=1}^{n-2} Z_i^\top  Q Z_i\right]^2\right\}\left(1+O_p\left(n^{-1/2}\right)\right)\\
		&=&\left\{\frac{1}{(n-2)^2}\left[\sum_{i,j=1}^{n-2}\left(Z_i^\top  Q Z_j\right)^2\right]-\frac{1}{(n-2)^3}\left[\sum_{i=1}^{n-2} Z_i^\top Q  Z_i\right]^2\right\}\left(1+O_p\left(n^{-1/2}\right)\right),
	\end{eqnarray*}
	where $Q= \Sigma^{\frac{1}{2}}\Lambda \Sigma^{\frac{1}{2}}$. To show that $\mbox{tr}(U_n^2)-\frac{1}{n-2}(\mbox{tr} U_n)^2$ is a ratio consistent estimator of $\mbox{tr} U^2$, we first show that the expectation of this estimator converges in probability to $\mbox{tr} U^2$ and then show that its variance converges to $o((trU^2)^2)$.
	
	When $p \to \infty$, $Z_i^\top Q Z_i \sim N (\mbox{tr} (Q), 2 \mbox{tr} (Q^2))$. By using the Gaussian distribution, it's straightforward to show that
	\begin{equation}
		E\left[\left( Z_i^\top  Q  Z_i\right)^2 \right]=2\mbox{tr} (Q^2)+\left(\mbox{tr} Q\right)^2.
	\end{equation}
	When $i\neq j$,
	\begin{equation}
		E\left( Z_i^\top Q Z_i Z_j^\top Q Z_j\right)=\left(\mbox{tr} Q\right)^2,
	\end{equation}
	and
	\begin{equation}
		E\left[\left( Z_i^\top  Q Z_j\right)^2 \right]=\mbox{tr}(Q^2).
	\end{equation}
	Hence,
	\begin{eqnarray}\label{eq: expectation estimator}
		\lefteqn{E\left[\mbox{tr}( U_n^2)-\frac{1}{n-2}(\mbox{tr} U_n)^2\right]}&& \nonumber\\
		&=&\frac{((n-2)^2+(n-2))\mbox{tr}(Q^2)+(\mbox{tr} Q)^2 (n-2)}{(n-2)^2}-\frac{2(n-2)\mbox{tr}(Q^2)+(n-2)^2(\mbox{tr} Q)^2}{(n-2)^3} \nonumber\\
		&=&\frac{(n-2)^2+(n-2)-2}{(n-2)^2}\mbox{tr} (Q^2)\to\mbox{tr} (Q^2)=\mbox{tr} (U^2).
	\end{eqnarray}
	
	To conclude that $\mbox{tr}( U_n^2)-\frac{1}{n-2}(\mbox{tr} U_n)^2$ is a consistent estimator of $\mbox{tr}(U^2)$, we show that its variance converges to zero in probability. To simplify the notation, we denote
	\begin{equation*}
		W_1=\left[\sum_{i,j=1}^{n-2}\left( Z_i^\top  Q  Z_j\right)^2\right],
		W_2=\left[\sum_{i=1}^{n-2} Z_i^\top  Q Z_i\right]^2,
	\end{equation*}
	and the variance 
	\begin{eqnarray}\label{eq: variance split}
		\lefteqn{\mbox{Var}\left[\mbox{tr}( U_n^2)-\frac{1}{n-2}(\mbox{tr} U_n)^2 \right]}&& \nonumber\\
		&=& \frac{1}{(n-2)^4}\mbox{Var} (W_1) + \frac{1}{(n-2)^6} \mbox{Var}(W_2) - \frac{1}{(n-2)^5}\mbox{Cov}(W_1, W_2)   .
	\end{eqnarray}
	Since the covariance $ - \sqrt{\mbox{Var}(W_1)}\sqrt{\mbox{Var}(W_2)} \le\mbox{Cov}(W_1, W_2) \le \sqrt{\mbox{Var}(W_1)}\sqrt{\mbox{Var}(W_2)}$, if we can show that the variance terms converge to zero in probability, the variance in \eqref{eq: variance split} converges to zero.
	
	From \cite[Lemma 2.1 and 2.2]{srivastava2009test}, it follows that 
	\begin{equation*}
		\mbox{Var}\left(Z_i^\top Q Z_i\right)=2\mbox{tr}(Q^2).
	\end{equation*}
	\begin{equation*}
		\mbox{Var}\left[\left(Z_i^\top Q Z_i\right)^2 \right]=8(\mbox{tr}Q)^2\mbox{tr}(Q^2)+8(\mbox{tr}Q^2)^2.
	\end{equation*} 
	When $i\neq j$,
	\begin{equation*}
		\mbox{Var}\left[\left(Z_i^\top Q Z_j\right)^2 \right]=6\mbox{tr} (Q^4)+2\left(\mbox{tr} (Q^2)\right)^2,
	\end{equation*}
	and
	\begin{equation*}
		\mbox{Var}\left(Z_i^\top  Q  Z_i Z_j^\top  Q Z_j\right)=4\left(\mbox{tr} (Q^2)\right)^2+4\mbox{tr} (Q^2)\left(\mbox{tr} (Q)\right)^2.
	\end{equation*}

	\begin{equation*}
		\mbox{Cov}\left(\left( Z_i^\top  Q  Z_i\right)^2,\left( Z_i^\top  Q Z_j\right)^2\right)\le \sqrt{8(\mbox{tr}Q)^2\mbox{tr}(Q^2)+8(\mbox{tr}Q^2)^2}\sqrt{6\mbox{tr}(Q^4)+2(\mbox{tr}( Q^2))^2},
	\end{equation*}
	
	and
	\begin{eqnarray*}
		\lefteqn{\mbox{Cov}\left(\left( Z_i^\top  Q  Z_i\right)^2,\left(Z_i^\top  Q  Z_i\right)\left( Z_j^\top  Q  Z_j\right)\right)} &&\\
		&=& E\left[\left( Z_i^\top Q  Z_i\right)^3\left( Z_j^\top  Q Z_j\right)\right]-E\left( Z_i^\top  Q  Z_i\right)^2E\left( Z_i^\top  Q Z_i\right)\left( Z_j^\top  Q Z_j\right) \\
		&=& E\left[\left( Z_i^\top Q  Z_i\right)^3\right ] E\left[\left( Z_j^\top  Q Z_j\right)\right]-E\left( Z_i^\top  Q  Z_i\right)^2E\left( Z_i^\top  Q Z_i\right)E\left( Z_j^\top  Q Z_j\right) \\
		&=& (6\mbox{tr} (Q^2)\mbox{tr} (Q)+(\mbox{tr} Q)^3)\mbox{tr}(Q) - (2\mbox{tr} (Q^2)+\left(\mbox{tr} Q\right)^2)\left(\mbox{tr} Q\right)^2  \\
		&=& 4\mbox{tr}(Q^2) (\mbox{tr} Q)^2,
	\end{eqnarray*}
	where the second equality holds by the independence between $Z_i$ and $Z_j$; the third equality holds by using $Z_i^\top Q Z_i \sim N (\mbox{tr} (Q), 2 \mbox{tr} (Q^2))$ when $p \to \infty$ and plug in the corresponding moments of the Gaussian distribution.
	
	When $i \neq j \neq k$,  
	\begin{eqnarray*}
		\lefteqn{\mbox{Cov}\left(\left( Z_i^\top  Q  Z_i\right)\left( Z_j^\top  Q  Z_j\right),\left(Z_i^\top  Q  Z_i\right)\left( Z_k^\top  Q  Z_k\right)\right)}&&\\
		&=&E\left[\left( Z_i^\top Q  Z_i\right)^2\left( Z_j^\top  Q Z_j\right)\left( Z_k^\top  Q Z_k\right)\right]-\left[E\left( Z_i^\top  Q  Z_i\right)\right]^2E\left( Z_j^\top  Q Z_j\right)E\left( Z_k^\top  Q Z_k\right)\\
		&=&2\mbox{tr}(Q^2) (\mbox{tr} Q)^2,
	\end{eqnarray*}
	and
	\begin{equation*}
		\mbox{Cov}\left(\left( Z_i^\top  Q  Z_j\right)^2,\left( Z_k^\top  Q Z_j\right)^2\right)\le 6\mbox{tr} (Q)^4+2\left(\mbox{tr} (Q)^2\right)^2.
	\end{equation*}

	Combining the above and by Assumption~\ref{cond: Q},
	\begin{eqnarray}\label{eq: variance of W1}
		\lefteqn{\frac{1}{(n-2)^4}\mbox{Var}(W_1)\le\frac{1}{(n-2)^4}\Big[(n-2)(8(\mbox{tr}Q)^2\mbox{tr}(Q^2)+8(\mbox{tr}Q^2)^2)}&&  \nonumber\\
		&&+2(n-2)(n-3)\left(6\mathrm{tr} (Q^4)+2\left(\mbox{tr} Q^2\right)^2\right)\Big]  \nonumber\\
		&&+\frac{1}{(n-2)^4}\left[4(n-2)(n-3)\sqrt{8(\mbox{tr}Q)^2\mbox{tr}(Q^2)+8(\mbox{tr}Q^2)^2}\sqrt{6\mbox{tr}(Q^4)+2(\mbox{tr}( Q^2))^2}\right] \nonumber\\
		&&+\frac{1}{(n-2)^4}\left[4(n-2)(n-3)(n-4)\left(6\mbox{tr} (Q)^4+2\left(\mbox{tr} (Q)^2\right)^2\right)\right] \nonumber\\
		&& = O(\mbox{tr} (Q^2)) =o\left((\mbox{tr} Q^2)^2\right) =o\left((\mbox{tr} U^2)^2\right),
	\end{eqnarray}
	\begin{eqnarray}\label{eq: variance of W2}
		\lefteqn{\frac{1}{(n-2)^6} \mbox{Var}(W_2)=\frac{1}{(n-2)^6}\Big[(n-2)(8(\mbox{tr}Q)^2\mbox{tr}(Q^2)+8(\mbox{tr}Q^2)^2)} &&  \nonumber\\
		&&+2(n-2)(n-3)\left(4\left(\mbox{tr} Q^2\right)^2+4\mbox{tr} (Q^2)\left(\mbox{tr}Q\right)^2\right)\Big] \nonumber\\
		&&+\frac{1}{(n-2)^6}\Big[4(n-2)(n-3)\mbox{tr} (Q^2)(\mbox{tr}Q)^2 +8(n-2)(n-3)(n-4)\mbox{tr}(Q^2) (\mbox{tr} Q)^2\Big]  \nonumber\\
		&& = o(\mbox{tr} (Q^2)) = o(\mbox{tr} (U^2)).
	\end{eqnarray}
	
	Hence, \eqref{eq: expectation estimator} suggests that
	\[
	E\left[\frac{\mbox{tr}(U_n^2) - \frac{1}{n-2} (\mbox{tr} U_n)^2}{\mbox{tr} (U^2)}\right] \to 1.
	\]
	Further, \eqref{eq: variance split}, \eqref{eq: variance of W1}, \eqref{eq: variance of W2} suggests that
	\[
	\mbox{Var}\left[\frac{\mbox{tr}(U_n^2) - \frac{1}{n-2} (\mbox{tr} U_n)^2}{\mbox{tr}(U^2)}\right] \to 0.
	\]
	
	Thus, by Lemma~\ref{lem: convergence in prob}, $\mbox{tr}( U_n^2)-\frac{1}{n-2}(\mbox{tr} U_n)^2$ is a ratio consistent estimator of $\mbox{tr} U^2$.
\end{enumerate}

\subsection{Proof of Lemma~\ref{lem: lambda replace}}\label{ssec: proof of lemma lambda replace}
We denote
\[
\Delta_n=\Lambda_n-\Lambda,
\qquad
E_n=n_0D^\top \Delta_n D-\operatorname{tr}(\Delta_n\Sigma).
\]
We show that
\[
E_n=o_P\left(\sqrt{\operatorname{tr}(U^2)}\right).
\]

Since $\Delta_n$ is diagonal, we may write
\[
\Delta_n=B_n\Lambda,
\]
where $B_n=\operatorname{diag}(b_{n1},\ldots,b_{np})$ with
\[
b_{nj}
=
\frac{[\Sigma]_{jj}-[S_n]_{jj}}{[S_n]_{jj}+k},
\qquad j=1,\ldots,p.
\]
Indeed, for each $j$,
\[
[\Delta_n]_{jj}
=
\frac{1}{[S_n]_{jj}+k}-\frac{1}{[\Sigma]_{jj}+k}
=
\frac{[\Sigma]_{jj}-[S_n]_{jj}}{[S_n]_{jj}+k}\cdot \frac{1}{[\Sigma]_{jj}+k}
=
b_{nj}[\Lambda]_{jj}.
\]

By Assumption~\ref{cond: bounded eigen value}, $[\Sigma]_{jj}+k$ is uniformly bounded away from $0$. Since the data are Gaussian and $p/n\to c_1\in(0,\infty)$, we have $\max_{1\le j\le p}|[S_n]_{jj}-[\Sigma]_{jj}|=o_P(1)$,
and therefore
\[
\|B_n\|=\max_{1\le j\le p}|b_{nj}|=o_P(1).
\]

Under Gaussian samples, $D$ and $S_n$ are independent. Thus, it holds that
$D \mid \Delta_n \sim N\left(\delta,\frac{\Sigma}{n_0}\right)$.
Therefore, it follows from the Gaussian quadratic form that 
\[
E(E_n\mid \Delta_n)=n_0\delta^\top \Delta_n\delta,
\]
and
\[
\operatorname{Var}(E_n\mid \Delta_n)
=
2\operatorname{tr}\bigl((\Delta_n\Sigma)^2\bigr)
+
4n_0\delta^\top \Delta_n\Sigma\Delta_n\delta.
\]

Now, since $\Delta_n=B_n\Lambda$,
\[
\operatorname{tr}\bigl((\Delta_n\Sigma)^2\bigr)
=
\operatorname{tr}\bigl((B_n\Lambda\Sigma)^2\bigr)
\le
\|B_n\|^2\,\operatorname{tr}\bigl((\Lambda\Sigma)^2\bigr)
=
\|B_n\|^2\,\operatorname{tr}(U^2)
=
o_P\bigl(\operatorname{tr}(U^2)\bigr).
\]
Similarly,
\[
n_0\delta^\top \Delta_n\Sigma\Delta_n\delta
=
n_0\delta^\top B_n\Lambda\Sigma\Lambda B_n\delta
\le
\|B_n\|^2\, n_0\delta^\top \Lambda\Sigma\Lambda\delta
=
o_P\bigl(\operatorname{tr}(U^2)\bigr),
\]
where the last step follows from Assumption~\ref{cond: negligible} together with $\|B_n\|=o_P(1)$.

Consequently,
\[
\frac{\operatorname{Var}(E_n\mid \Delta_n)}{\operatorname{tr}(U^2)}
\stackrel{P}{\to}0.
\]
By Chebyshev's inequality, this implies
\[
\frac{E_n-E(E_n\mid \Delta_n)}{\sqrt{2\operatorname{tr}(U^2)}}
\stackrel{P}{\to}0.
\]

Under $H_0$, we have $\delta=0$, and hence
\[
E(E_n\mid \Delta_n)=0.
\]
Therefore,
\[
\frac{E_n}{\sqrt{2\operatorname{tr}(U^2)}}
\stackrel{P}{\to}0.
\]

Under the alternatives, the same conclusion remains valid provided that
\[
n_0\delta^\top \Delta_n\delta
=
o_P\left(\sqrt{\operatorname{tr}(U^2)}\right).
\]
This completes the proof.

\subsection{Proof of Lemma~\ref{prop: remainder converge to zero}}\label{ssec: proof of prop 2}
From the proof of Proposition~\ref{prop: consistent estimators} and Assumption~\ref{cond: Q}, we only need to prove 
\begin{equation}\label{eq: converge to zero}
	\frac{\mathrm{tr}(\Lambda(S_n-\Sigma))}{\sqrt{2\mathrm{tr}(U^2)}}\to0.
\end{equation}
Rewrite as
\begin{equation}
	\mathcal{G}=\frac{\frac{1}{n-2}\sum\limits_{i=1}^{n-2}Z_i^\top   QZ_i-\mathrm{tr}(\Lambda\Sigma)}{\sqrt{2\mathrm{tr}(U^2)}}\to0,
\end{equation}
By Assumption~\ref{cond: bounded eigen value}, it holds that 
\begin{align*}
	E(\mathcal{G})&=0\\
	Var(\mathcal{G})&=\frac{\frac{1}{n-2}\mathrm{tr}(Q^2)}{2\mathrm{tr}(U^2)}\to0.
\end{align*}
This completes the proof.
\subsection{Proof of Theorem~\ref{thm: asymptotic normality}}\label{ssec: proof of theorem 1}
\begin{enumerate}
	\item Under $H_0$, with (\ref{eq: test stat split2}), we only need to show
	$$\dfrac{n_0 D^\top \Lambda D-\mbox{tr}(U)}{\sqrt{2\mbox{tr}(U^2)}}\stackrel{d}{\to} N(0,1).$$
	Note that
	$$\theta =\sqrt{n_0} \Sigma^{-\frac{1}{2}} D\sim N_p( 0,{I}_p).$$
	Hence, 
	\begin{equation}\label{eq: population test stats}
		\dfrac{n_0 D^\top \Lambda D-\mbox{tr}(U)}{\sqrt{2\mbox{tr}(U^2)}} =\frac{{\theta}^\top \Sigma^{\frac{1}{2}} \Lambda \Sigma^{\frac{1}{2}}\theta-\mbox{tr}( U)}{\sqrt{2\mbox{tr}( U^2)}}.
	\end{equation}
	Further, we rewrite \eqref{eq: population test stats} by an eigen decomposition on $\Sigma^{\frac{1}{2}} \Lambda \Sigma^{\frac{1}{2}}$ to $\Phi\Gamma \Phi^\top$,
	where $\Gamma$ is a diagonal matrix with the elements $\gamma_1,\cdots,\gamma_p$. From the eigen decomposition, it's easy to obtain 
	\[\mbox{tr}(\Sigma^{\frac{1}{2}} \Lambda \Sigma^{\frac{1}{2}})=\mbox{tr}(\Gamma)=\mbox{tr}(U).\] 
	
	Let $\beta= \Phi^\top \theta\sim N_p(0,{I}_p)$, 
	\begin{equation}\label{eq: population test rewrite}
		\frac{\beta^\top \Gamma \beta-\mbox{tr}(\Gamma)}{\sqrt{2\mbox{tr}( U^2)}}=\frac{\sum_{i=1}^p\gamma_i(\beta_i^2-1)}{\sqrt{2\mbox{tr}( U^2)}}.
	\end{equation}
	By Lemma~\ref{lem: moments of standard Gaussian}, $\beta_j \sim N(0,1)$ and $\tau_i=\gamma_i(\beta_i^2-1)$, it's straightforward to show that
	$${E}\tau_i=0,{E}\tau_i^2=2\gamma_i^2, {E}\tau_i^4=60\gamma_i^4.$$
	
	Further, since $\gamma_i$'s are the eigenvalues of $\Sigma^{\frac{1}{2}}\Lambda \Sigma^{\frac{1}{2}}$, it follows that $\gamma_i^2$'s are the eigenvalues of $\Sigma^{\frac{1}{2}}\Lambda \Sigma\Lambda\Sigma^{\frac{1}{2}}$. 
	
	Notice that
	$$\sum_{i=1}^p2\gamma_i^2 = 2\mbox{tr}(\Sigma^{\frac{1}{2}}\Lambda \Sigma \Lambda\Sigma^{\frac{1}{2}})=2\mbox{tr}(\Lambda \Sigma)^2=2\mbox{tr}(U^2),$$
	and similarly,
	$$\sum_{i=1}^p\gamma_i^4=\mbox{tr}( U^4).$$
	By taking $B_n^2=\sum_{i=1}^p2\gamma_i^2 = 2 \mbox{tr}(U^2)$ and $\delta = 2$ in Lemma~\ref{lem: convergence} and combining Assumption~\ref{A1}, the following convergence holds 
	$$  \dfrac{n_0 D^\top \Lambda D-\mbox{tr}(U)}{\sqrt{2\mbox{tr}(U^2)}} \stackrel{d}{\to} N(0,1).$$
	\item Under $H_1$,
	$\theta\sim N(\sqrt{n_0} \Sigma^{-\frac{1}{2}} \delta,I_p)$, $\theta_0= \theta-\sqrt{n_0} \Sigma^{-\frac{1}{2}} \delta, \theta_0\sim N_p(0, I_p)$, by \eqref{eq: population test stats} and \eqref{eq: population test rewrite}, 
	\begin{eqnarray}\label{eq: alternative test stats decomposition}
		\sum_{i=1}^p \dfrac{\gamma_i(\beta_i^2-1)}{\sqrt{2\mbox{tr}( U^2)}}=\dfrac{\theta_0^\top \Sigma^{\frac{1}{2}} \Lambda \Sigma^{\frac{1}{2}} \theta_0-\mbox{tr}(\Gamma)}{\sqrt{2\mbox{tr}(U^2)}} +\dfrac{n_{0}\delta^\top \Lambda \delta}{\sqrt{2\mbox{tr}( U^2)}}+2\dfrac{\sqrt{n_{0}}\delta^\top \Lambda \Sigma^{\frac{1}{2}} \theta_0}{\sqrt{2\mbox{tr}(U^2)}}.
	\end{eqnarray}
	The last term of \eqref{eq: alternative test stats decomposition} is a remainder term and we denote it as $\mathcal{R}=\dfrac{\sqrt{n_{0}} \delta^\top \Lambda \Sigma^{\frac{1}{2}}\theta_0}{\sqrt{2\mbox{tr}( U^2)}}$. By Assumption~\ref{cond: negligible}, it is easy to obtain
	$$ E (\mathcal{R})=0, \mbox{Var}(\mathcal{R})=\frac{n_0 \delta^\top \Lambda \Sigma \Lambda\delta}{2\mbox{tr}(U^2)} = o(1),$$
	which suggests that $ \mathcal{R}\stackrel{P}{\to} 0$ by Lemma~\ref{lem: convergence in prob}. Since the first term of \eqref{eq: alternative test stats decomposition} tends to standard normal, we can conclude that
	$$\frac{n_{0} D^\top \Lambda  D-\mbox{tr}( U)-n_{0} \delta^\top \Lambda \delta}{\sqrt{2\mbox{tr}( U^2)}}\stackrel{d}{\to} N(0,1),$$
	with \eqref{eq: test stat split} and Proposition 2,\;rewrite
	$$\frac{n_{0} D^\top \Lambda_n  D-\mbox{tr}( U_n)-n_{0} \delta^\top \Lambda \delta}{\sqrt{2\mbox{tr}( U^2)}}\stackrel{d}{\to} N(0,1).$$
\end{enumerate}

\subsection{Proof of Corollary~\ref{Cor: power of the test}}
Notice that
\begin{eqnarray*}
	\lefteqn{P(T_{\rm{BF}}\ge u_{1-\alpha}) }&&\\
	&=&P\left(\dfrac{n_0 D^\top \Lambda_n  D-\mbox{tr}( U_n)-n_0\delta^\top \Lambda \delta}{\sqrt{2\mbox{tr}(U^2)}}\ge u_{1-\alpha}-\frac{n_0 \delta^\top \Lambda \delta}{\sqrt{2\mbox{tr}( U^2)}}\right)\\
	&=& P\left(\dfrac{-n_0 D^\top \Lambda_n D+\mbox{tr}(U_n)+n_0 \delta^\top \Lambda\delta}{\sqrt{2\mbox{tr}(U^2)}}\le -u_{1-\alpha}+\frac{n_0 \delta^\top \Lambda \delta}{\sqrt{2\mbox{tr}(U^2)}}\right),
\end{eqnarray*}
thus,
$$g(\delta)-\Phi\left(-u_{1-\alpha}+\frac{n_{0}\delta^\top \Lambda\delta}{\sqrt{2{\mbox{tr}(U^2)}}}\right)\to0,$$
where $g(\delta)$ is the power of the $T_{\rm{BF}}$ test.\\

\section{Simulation tables}\label{section: simulation tables}

\begin{table}[htpb!]
	\centering
	\setlength{\tabcolsep}{10mm}
	\renewcommand{\arraystretch}{1}
	\begin{tabular}{lcccc}
		\toprule
		Test & $\Sigma_1$ & $\Sigma_2$ & $\Sigma_3$ & $\Sigma_4$ \\
		\midrule
		BF  & 0.0550 & 0.0584 & 0.0520 & 0.0528 \\
		SD  & 0.0504 & 0.0486 & 0.0516 & 0.0448 \\
		PB  & 0.0576 & 0.0584 & 0.0586 & 0.0596 \\
		BS  & 0.0568 & 0.0618 & 0.0602 & 0.0656 \\
		CQ  & 0.0520 & 0.0594 & 0.0604 & 0.0618 \\
		RM  & 0.0438 & 0.0502 & 0.0558 & 0.0386 \\
		\bottomrule
	\end{tabular}
	\caption{The sizes of six tests under different covariance matrix specifications.}
	\label{table:test_sizes}
\end{table}

\vspace{-3em}

\begin{table}[ht!]
	\centering
	\begin{tabular}{c c c c c c| c c c c c}
		\toprule
		&\multicolumn{5}{c}{Alternative 1}&\multicolumn{5}{c}{Alternative 2}\\
		\midrule
		$p_0$&0.5&0.6&0.7&0.8&0.9&0.5&0.6&0.7&0.8&0.9\\
		\hline
		BF&0.7458&0.7534&0.7494&0.7536&0.7532&0.5276&0.5340&0.5216&0.5290&0.5232\\
		\hline
		PB&0.7418&0.7504&0.7402&0.7520&0.7458&0.5234&0.5334&0.5372&0.5282&0.5304\\
		\hline
		SD&0.7152&0.7120&0.7094&0.7034&0.7154&0.4906&0.4878&0.5022&0.4876&0.4964\\
		\hline
		BS&0.7440&0.7414&0.7420&0.7370&0.7446&0.5272&0.5224&0.5276&0.5268&0.5376\\
		\hline
		
		CQ&0.7526&0.7422&0.7410&0.7540&0.7544&0.5236&0.5308&0.5334&0.5232&0.5354\\
		\hline
		
		RM&0.6638&0.6420&0.6558&0.6324&0.6610&0.4632&0.4780&0.4726&0.4768&0.4750\\
		\bottomrule
	\end{tabular}
	\caption{The powers of six tests under $\Sigma_1$.}
\end{table}

\begin{table}[ht!]
	\centering
	\begin{tabular}{c c c c c c| c c c c c}
		\toprule
		&\multicolumn{5}{c}{Alternative 1}&\multicolumn{5}{c}{Alternative 2}\\
		\midrule
		$p_0$&0.5&0.6&0.7&0.8&0.9&0.5&0.6&0.7&0.8&0.9\\
		\hline
		BF&0.5164&0.5194&0.4648&0.4824&0.4416&0.5144&0.5290&0.5238&0.5284&0.5232\\
		\hline
		PB&0.5342&0.5048&0.4952&0.5284&0.5046&0.5224&0.5302&0.5278&0.5366&0.5224\\
		\hline
		SD&0.5118&0.4778&0.4802&0.4718&0.4090&0.4694&0.4704&0.4692&0.4862&0.4744\\
		\hline
		BS&0.5228&0.5354&0.5112&0.5238&0.4488&0.5212&0.5206&0.5004&0.5140&0.5164\\
		\hline
		
		CQ&0.5282&0.5362&0.5008&0.4738&0.4948&0.5200&0.5334&0.5202&0.5300&0.5210\\
		\hline
		
		RM&0.5314&0.4672&0.4784&0.4470&0.4908&0.4988&0.5110&0.4966&0.4828&0.5250\\
		\bottomrule
	\end{tabular}
	\caption{The powers of six tests under $\Sigma_2$.}
\end{table}

\begin{table}[ht!]
	\centering
	\begin{tabular}{c c c c c c| c c c c c}
		\toprule
		&\multicolumn{5}{c}{Alternative 1}&\multicolumn{5}{c}{Alternative 2}\\
		\midrule
		$p_0$&0.5&0.6&0.7&0.8&0.9&0.5&0.6&0.7&0.8&0.9\\
		\hline
		BF&0.6012&0.6448&0.5976&0.5776&0.5652&0.6966&0.7046&0.6794&0.6876&0.7292\\
		\hline
		PB&0.4290&0.4094&0.3894&0.3994&0.3680&0.5102&0.5126&0.5192&0.5020&0.5064\\
		\hline
		SD&0.5882&0.6030&0.6458&0.5358&0.5274&0.6740&0.6482&0.6798&0.7016&0.6306\\
		\hline
		BS&0.4626&0.4196&0.4062&0.4142&0.3762&0.5064&0.4972&0.5074&0.5098&0.5168\\
		\hline
		
		CQ&0.4908&0.4370&0.4254&0.4182&0.3786&0.5162&0.5154&0.5148&0.5046&0.5174\\
		\hline
		
		RM&0.5676&0.5538&0.5212&0.5456&0.4976&0.6712&0.6748&0.6736&0.5980&0.6458\\
		\bottomrule
	\end{tabular}
	\caption{The powers of six tests under $\Sigma_3$.}
\end{table}

\begin{table}[ht!]
	\centering
	\begin{tabular}{c c c c c c| c c c c c}
		\toprule
		&\multicolumn{5}{c}{Alternative 1}&\multicolumn{5}{c}{Alternative 2}\\
		\midrule
		$p_0$&0.5&0.6&0.7&0.8&0.9&0.5&0.6&0.7&0.8&0.9\\
		\hline
		BF&0.7208&0.7398&0.7146&0.7302&0.7336&0.8944&0.9018&0.9132&0.8906&0.8566\\
		\hline
		PB&0.3170&0.3155&0.3580&0.3570&0.3250&0.5025&0.4875&0.5070&0.5005&0.4925\\
		\hline
		SD&0.7090&0.7180&0.7124&0.7198&0.7078&0.9046&0.9114&0.8812&0.9096&0.8984\\
		\hline
		BS&0.3096&0.3280&0.3440&0.3074&0.3108&0.5086&0.4984&0.4968&0.5044&0.4988\\
		\hline
		
		CQ&0.3186&0.3192&0.3418&0.3442&0.3130&0.5092&0.4992&0.4816&0.5112&0.4920\\
		\hline
		
		RM&0.5962&0.5908&0.5612&0.5850&0.6032&0.5852&0.6328&0.5728&0.6082&0.6826\\
		\bottomrule
	\end{tabular}
	\caption{The powers of six tests under $\Sigma_4$.}
\end{table}

\begin{table}[ht!]
	\centering
	\begin{tabular}{c c c c c c| c c c c c}
		\toprule
		&\multicolumn{5}{c}{Size}&\multicolumn{5}{c}{Power}\\
		\midrule
		$n_1=n_2$&20&30&40&50&60&20&30&40&50&60\\
		\hline
		BF&0.0394&0.0470&0.0514&0.0462&0.0496&0.7974&0.9646&0.9970&0.9846&1\\
		\hline
		PB&0.0598&0.060&0.0562&0.0552&0.0638&0.4370&0.7028&0.8926&0.9220&0.9954\\
		\hline
		SD&0.0332&0.0352&0.0402&0.0428&0.0434&0.7732&0.9562&0.9958&0.9996&1\\
		\hline
		BS&0.0548&0.0662&0.0586&0.0602&0.0662&0.4266&0.7008&0.8998&0.9760&0.9950\\
		\hline
		CQ&0.0568&0.0618&0.0584&0.0574&0.0606&0.4340&0.7074&0.8978&0.9710&0.9960\\
		\hline
		RM&0.0476&0.0482&0.0492&0.0438&0.0396&0.6924&0.9324&0.9824&0.9992&1\\
		\bottomrule
	\end{tabular}
	\caption{The sizes and powers of the tests for misspecified distributions under $\Sigma_4$.}
	\label{table: misspecified Gaussian sigma 4}
\end{table}

\begin{table}[ht!]
	\centering
	\begin{tabular}{c c c c c c| c c c c c}
		\hline
		&\multicolumn{5}{c}{Size}&\multicolumn{5}{c}{Power}\\
		\hline
		$n_1=n_2$&20&30&40&50&60&20&30&40&50&60\\
		\hline
		BF&0.0502&0.0522&0.0518&0.0554&0.0502&0.1964&0.2910&0.3994&0.5354&0.6960\\
		\hline
		PB&0.0494&0.0564&0.0516&0.0524&0.0516&0.1410&0.2238&0.3148&0.4386&0.5866\\
		\hline
		SD&0.0412&0.0382&0.0392&0.0446&0.0442&0.1450&0.2756&0.3854&0.5196&0.7498\\
		\hline
		BS&0.0534&0.0524&0.0536&0.0516&0.0464&0.1406&0.2136&0.3364&0.4852&0.5732\\
		\hline
		CQ&0.0502&0.0496&0.0462&0.0546&0.0484&0.1410&0.2374&0.3408&0.4484&0.5852\\
		\hline
		RM&0.0396&0.0502&0.0564&0.0496&0.0456&0.1408&0.2622&0.4318&0.6812&0.8322\\
		\hline
	\end{tabular}	
	\caption{The sizes and powers of the tests for misspecified distributions under $\Sigma_5$.}
	\label{table: misspecified Gaussian sigma 5}
\end{table}

\end{appendices}

\clearpage

\bibliography{BayesFactor}

\begin{thebibliography}{23}
\providecommand{\natexlab}[1]{#1}
\providecommand{\url}[1]{{#1}}
\providecommand{\urlprefix}{URL }
\providecommand{\doi}[1]{\url{https://doi.org/#1}}
\providecommand{\eprint}[2][]{\url{#2}}
 \bibcommenthead

\bibitem[{Aitkin(1991)}]{aitkin1991posterior}
Aitkin M (1991) Posterior {Bayes} factors. Journal of the Royal Statistical
  Society Series B: Statistical Methodology 53(1):111--128

\bibitem[{Anderson(1958)}]{anderson1958introduction}
Anderson TW (1958) An Introduction to Multivariate Statistical Analysis. Wiley,
  New York

\bibitem[{Bai and Saranadasa(1996)}]{bai1996effect}
Bai Z, Saranadasa H (1996) Effect of high dimension: By an example of a two
  sample problem. Statistica Sinica 6:311--329

\bibitem[{Cai et~al.(2014)Cai, Liu, and Xia}]{tony2014two}
Cai T, Liu W, Xia Y (2014) Two-sample test of high dimensional means under
  dependence. Journal of the Royal Statistical Society Series B: Statistical
  Methodology 76(2):349--372

\bibitem[{Chen et~al.(2024)Chen, Hai, and Wang}]{chen2024bayesian}
Chen F, Hai Q, Wang M (2024) {Bayesian} hypothesis testing for equality of
  high-dimensional means using cluster subspaces. Computational Statistics
  39(3):1301--1320

\bibitem[{Chen and Qin(2010)}]{chen2010two}
Chen S, Qin Y (2010) A two-sample test for high-dimensional data with
  applications to gene-set testing. The Annals of Statistics 38(2):808--835

\bibitem[{Guhaniyogi and Dunson(2015)}]{guhaniyogi2015bayesian}
Guhaniyogi R, Dunson DB (2015) {Bayesian} compressed regression. Journal of the
  American Statistical Association 110(512):1500--1514

\bibitem[{Hotelling(1931)}]{hotelling1931generalization}
Hotelling H (1931) The generalization of {Student's} ratio. Annals of
  Mathematical Statistics 2:360--378

\bibitem[{Huang et~al.(2022)Huang, Li, Li, and Yang}]{huang2022overview}
Huang Y, Li C, Li R, et~al (2022) An overview of tests on high-dimensional
  means. Journal of Multivariate Analysis 188:104813

\bibitem[{Jiang and Xu(2022)}]{jiang2022two}
Jiang Y, Xu X (2022) A two-sample test of high dimensional means based on
  posterior {Bayes} factor. Mathematics 10(10):1741

\bibitem[{Lee et~al.(2024)Lee, You, and Lin}]{lee2024bayesian}
Lee K, You K, Lin L (2024) {Bayesian} optimal two-sample tests for
  high-dimensional {Gaussian} populations. Bayesian Analysis 19(3):869--893

\bibitem[{Lopes et~al.(2011)Lopes, Jacob, and Wainwright}]{lopes2011more}
Lopes M, Jacob L, Wainwright M (2011) A more powerful two-sample test in high
  dimensions using random projection. Advances in Neural Information Processing
  Systems 24:1206--1214

\bibitem[{Shiryaev(2016)}]{shiryaev2016probability}
Shiryaev A (2016) Probability, Graduate Texts in Mathematics, vol~95, 2nd edn.
  Springer, New York

\bibitem[{Srivastava(2009)}]{srivastava2009test}
Srivastava MS (2009) A test for the mean vector with fewer observations than
  the dimension under non-normality. Journal of Multivariate Analysis
  100(3):518--532

\bibitem[{Srivastava and Du(2008)}]{srivastava2008test}
Srivastava MS, Du M (2008) A test for the mean vector with fewer observations
  than the dimension. Journal of Multivariate Analysis 99(3):386--402

\bibitem[{Srivastava et~al.(2013)Srivastava, Katayama, and
  Kano}]{srivastava2013two}
Srivastava MS, Katayama S, Kano Y (2013) A two sample test in high dimensional
  data. Journal of Multivariate Analysis 114:349--358

\bibitem[{Srivastava et~al.(2016)Srivastava, Li, and
  Ruppert}]{srivastava2016raptt}
Srivastava R, Li P, Ruppert D (2016) {RAPTT}: An exact two-sample test in high
  dimensions using random projections. Journal of Computational and Graphical
  Statistics 25(3):954--970

\bibitem[{Thulin(2014)}]{thulin2014high}
Thulin M (2014) A high-dimensional two-sample test for the mean using random
  subspaces. Computational Statistics \& Data Analysis 74:26--38

\bibitem[{Wang et~al.(2024)Wang, Ye, and Han}]{wang2024bayesian}
Wang M, Ye K, Han Z (2024) {Bayesian} analysis of testing general hypotheses in
  linear models with spherically symmetric errors. Test 33(1):251--270

\bibitem[{Wang and Xu(2021)}]{wang2021bayesian}
Wang R, Xu X (2021) A {Bayesian}-motivated test for high-dimensional linear
  regression models with fixed design matrix. Statistical Papers
  62(4):1821--1852

\bibitem[{Xue and Yao(2020)}]{xue2020}
Xue K, Yao F (2020) Distribution and correlation-free two-sample test of
  high-dimensional means. The Annals of Statistics 48(3):1304--1328

\bibitem[{Yang et~al.(2024)Yang, Zheng, and Li}]{yang2024new}
Yang S, Zheng S, Li R (2024) A new test for high-dimensional two-sample mean
  problems with consideration of correlation structure. The Annals of
  Statistics 52(5):2217--2240

\bibitem[{Zoh et~al.(2018)Zoh, Sarkar, Carroll, and Mallick}]{zoh2018powerful}
Zoh R, Sarkar A, Carroll RJ, et~al (2018) A powerful {Bayesian} test for
  equality of means in high dimensions. Journal of the American Statistical
  Association 113(524):1733--1741

\end{thebibliography}

\end{document}